\newif\ifdraft
\newif\iffull
\newif\ifcomment
\newif\iflatexdiff
\newif\ifbibtex
\newif\ifpreprint
\newif\ifbanner
\newif\ifsupp
\latexdifftrue   
\bibtextrue      
\preprinttrue    
\fulltrue        

\def\dvers{v8}

\ifpreprint

\fi
\def\dtitle{Multipion Bose-Einstein correlations \\
in pp, p--Pb, and Pb--Pb collisions at the LHC} 
\def\stitle{Multipion Bose-Einstein correlations} 
\ifpreprint
\documentclass[ALICE,manyauthors,12pt]{cernphprep}
\usepackage[comma,square,numbers,sort&compress]{natbib}
\usepackage{hyperref}
\else
\documentclass[final,3p,12pt]{elsarticle}
\biboptions{comma,square,numbers,sort&compress}
\usepackage{hyperref}
\fi
\usepackage{graphicx}  
\usepackage{dcolumn}   
\usepackage{bm}        
\usepackage{amssymb}   
\usepackage{amsfonts}
\usepackage{graphics}
\usepackage{grffile}   
\usepackage{epsfig}
\usepackage{units}
\usepackage[usenames]{color}
\usepackage[normalem]{ulem} 
\usepackage{color}
\usepackage[utf8]{inputenc}
\usepackage[T1]{fontenc}
\usepackage{subfigure}
\usepackage{verbatim}
\iflatexdiff
\RequirePackage{color}\definecolor{RED}{rgb}{1,0,0}\definecolor{BLUE}{rgb}{0,0,1}

\fi

\ifpreprint
\else
 
\fi

\newcommand{\kT}           {\ensuremath{K_{\rm T2}}}
\newcommand{\KTThree}      {\ensuremath{K_{\rm T3}}}
\newcommand{\KTFour}      {\ensuremath{K_{\rm T4}}}
\newcommand{\qinv}         {\ensuremath{q}}

\newcommand{\dEdx}         {d$E$/d$x$}

\newcommand{\pp}           {pp}

\newcommand{\PbPb}         {\mbox{Pb--Pb}}

\newcommand{\pPb}          {\mbox{p--Pb}}

\newcommand{\pt}           {\ensuremath{p_{\mathrm{T}}}{ }}

\newcommand{\snn}          {\ensuremath{\sqrt{s_{\mathrm{NN}}}}}

\newcommand{\red}[1]       {\textcolor{red}{#1}}

\newcommand{\warn}[1]      {{\small\textbf{\red{(!}\footnote{\textbf{\red{(!)}}~#1}\red{)}}}\marginpar{\textbf{\red{---}}}}

\newcommand{\com}[1]       {}

\iflatexdiff

\renewcommand{\xout}[1]    {\textcolor{red}{\sout{#1}}}
 
\newcommand{\old}[1]       {{\textcolor{red}{\sout{#1}}}}

\else

\renewcommand{\xout}[1]    {}
\newcommand{\old}[1]       {\relax}

\fi

\graphicspath{{./img/}}
\ifdraft
\usepackage{lineno}
\linenumbers
\modulolinenumbers[1]
\usepackage{fancyhdr}
\ifbanner
\fi

\else
\renewcommand{\warn}[1]{}

\fi
\begin{document}
\newlength{\figlen}
\setlength{\figlen}{\linewidth}
\ifpreprint
\setlength{\figlen}{0.95\linewidth}
\begin{titlepage}
\PHyear{2015}
\PHnumber{330}                        
\PHdate{22 December}                     
\title{\dtitle}
\ShortTitle{\stitle}
\Collaboration{ALICE Collaboration%
         \thanks{See Appendix~\ref{app:collab} for the list of collaboration members}}
\ShortAuthor{ALICE Collaboration} 
\ifdraft
\begin{center}
\ifbanner
 \today\\ \color{red}DRAFT \dvers\ \hspace{0.3cm} \$Revision: 1777 $\color{white}:$\$\color{black}\vspace{0.3cm}
\else
 \today\\ \color{red}DRAFT \dvers\ \color{black}\vspace{0.3cm}
\fi
\end{center}
\fi
\else
\begin{frontmatter}
\title{\dtitle}
\iffull
\input{Alice_Authorlist_2014-Mar-21-PLB}
\else
\ifdraft
\author{ALICE Collaboration \\ \vspace{0.3cm} 
\today\\ \color{red}DRAFT \dvers\ \hspace{0.3cm} (\$Revision: 1777 $\color{white}:$\$)\color{black}}
\else
\author{ALICE Collaboration}
\fi
\fi
\fi
\begin{abstract}
Three- and four-pion Bose-Einstein correlations are presented in pp, p--Pb, and Pb--Pb collisions at the LHC.
We compare our measured four-pion correlations to the expectation derived from two- and three-pion measurements.
Such a comparison provides a method to search for coherent pion emission.
We also present mixed-charge correlations in order to demonstrate the effectiveness of several analysis procedures such as Coulomb corrections.
Same-charge four-pion correlations in pp and p--Pb appear consistent with the expectations from three-pion measurements.  
However, the presence of non-negligible background correlations in both systems prevent a conclusive statement.
In Pb--Pb collisions, we observe a significant suppression of three- and four-pion Bose-Einstein correlations compared to expectations from two-pion measurements.
There appears to be no centrality dependence of the suppression within the 0--50\% centrality interval.  
The origin of the suppression is not clear.  However, by postulating either coherent pion emission or large multibody Coulomb effects, the suppression may be explained.

\ifdraft 
\ifpreprint
\end{abstract}
\end{titlepage}
\else
\end{abstract}
\end{frontmatter}
\newpage
\fi
\fi
\ifdraft
\thispagestyle{fancyplain}
\else
\end{abstract}
\ifpreprint
\end{titlepage}
\else
\end{frontmatter}
\fi
\fi
\setcounter{page}{2}


\section{Introduction}
\label{sec:intro}
The last stage of particle interactions in high-energy collisions (kinetic freeze-out) occurs on the femtoscopic length scale ($10^{-15}$ m) where quantum statistical (QS) correlations are expected. 
QS correlations at low relative momentum are known to be sensitive to the space-time extent (e.g.\ radius) and dynamics of the particle emitting source \cite{Brown:1956zza,Goldhaber:1960sf,Kopylov:1975rp}.
Another interesting, although less studied, aspect of QS correlations is the possible suppression due to coherent pion emission \cite{Gyulassy:1979yi,Andreev:1992pu,Plumer:1992au,Akkelin:2001nd}. 
Coherent emission may arise for several reasons such as from the formation of a disoriented chiral condensate (DCC) \cite{Bjorken:1993cz,Greiner:1993jn,Bjorken:1997re,Rajagopal:1997au}, gluonic or pionic Bose-Einstein Condensates (BEC) \cite{Ornik:1993gb,Blaizot:2011xf,Blaizot:2012qd,Begun:2015ifa}, or multiple coherent sources from pulsed radiation \cite{Ikonen:2008zz}.

Coherent emission is known to suppress Bose-Einstein correlations below the expectation from a fully chaotic particle emitting source. 
Some of the earliest attempts to search for coherence relied solely on fits to two-pion correlation functions \cite{Neumeister:1991bq}.  
The intercepts of the fits at zero relative momentum were found to be highly suppressed.
However, it was quickly realized that Coulomb repulsion and long-lived emitters (e.g.\ long-lived resonance decays) also suppress the correlation function significantly.
Furthermore, the precise shape of the freeze-out space-time distribution is unknown.  As a consequence, the corresponding functional form of the correlation function in momentum space is also unknown.  
Being such, there is no reliable way to extrapolate the measured correlation function to the unmeasured intercept.

Multipion Bose-Einstein correlations could provide an increased sensitivity to coherence as the expected suppression increases with the order of the correlation function \cite{Andreev:1992pu,Csorgo:1999sj,Gangadharan:2015ina}.
However, the analysis of multipion Bose-Einstein correlations comes at the expense of increased complexity.
Some of the earliest attempts to measure three-pion Bose-Einstein correlations relied on a different methodology and gave rather ambiguous results \cite{Boggild:1999tu,Aggarwal:2000ex,Bearden:2001ea,Adams:2003vd}.
Recently the methodology of isolating three- and four-pion Bose-Einstein correlations has been considerably improved \cite{Gangadharan:2015ina}--particularly in regards to the treatment of long-lived pion emitters.
Our previous measurements of three-pion correlations revealed a suppression which may arise from a coherent fraction ($G$) of $23\% \pm 8\%$ at low \pt~at kinetic freeze-out \cite{Abelev:2013pqa}.

We present three- and four-pion QS correlations in pp, p--Pb, and Pb--Pb collisions at the LHC measured with ALICE using the methodology presented in Ref.~\cite{Gangadharan:2015ina}.
The QS correlations are extracted from the measured multipion distributions.
The extraction of QS correlations relies on the treatment of long-lived pion emitters and final-state interactions (FSI), e.g.\ Coulomb correlations.
QS correlations between pions separated by large distances ($>\sim100$ fm) are only observable at very low relative momentum, where track merging effects and finite momentum resolution prevent reliable measurements.
The effect of long-lived emitters at measurable relative momentum is to simply dilute the correlation functions.
The presented correlation functions are corrected for this dilution as well as FSI and therefore should represent the pure QS correlations from short-lived pion emitters, i.e.\ the core of particle production.
We also present the mixed-charge four-pion correlations, which are used to demonstrate the effectiveness of all corrections in the analysis procedure.

The measured multipion QS correlations require a reference in order to quantify a possible suppression.
Lower order QS correlation functions form the reference in this analysis.
Two-pion QS correlations, in particular, provide a direct measurement of the pair-exchange magnitudes, which may be used as a building block to form an expectation for higher order correlation functions.
These ``expected'' multibody correlations were termed ``built'' in Ref.~ \cite{Gangadharan:2015ina}.

This article is organized into 7 sections.  
We explain the detector setup and data selection in Sec.~$2$.
In Sec.~$3$, we describe the analysis methodology.
The results are presented in Sec.~$4$.
In Sec.~$5$, we discuss all of the systematic uncertainties investigated.
We discuss several possible origins of the suppression in Sec.~$6$.
Finally, in Sec.~$7$ we summarize our findings.

\section{Experimental setup and data selection}
\label{sec:setup}
Data from \pp, \pPb\, and \PbPb\ collisions at the LHC recorded with ALICE~\cite{Aamodt:2008zz} are analyzed.   
The data for \pp\ collisions at $\sqrt{s}=7$~TeV were taken during 2010, during 2013 for \pPb\ collisions at $\snn=5.02$~TeV, and during 2011 for \PbPb\ at $\snn=2.76$~TeV.

The trigger conditions are slightly different for each of the three collision systems.
For pp collisions, at least one hit in the Silicon Pixel Detector (SPD), at central rapidity, or either of the V0 detectors~\cite{Abelev:2013qoq}, at forward rapidity, is required. For \PbPb\ and \pPb\ collisions, the trigger is formed by requiring hits simultaneously in each V0.

The Inner Tracking System (ITS) and Time Projection Chamber (TPC) located at mid-rapidity are used for particle tracking~\cite{Alme:2010ke}.
There are 6 layers of silicon detectors in the ITS: two silicon pixel, two silicon drift, and two silicon strip detectors.
The ITS provides high spatial resolution for the position of the primary vertex.
The TPC alone is used for momentum and charge determination of particles through the radius of curvature of the particles traversing a 0.5~T longitudinal magnetic field.
The TPC additionally provides particle identification through the specific ionization energy loss (\dEdx).  
To ensure uniform tracking, the $z$-coordinate (along the beam-axis) of the primary vertex 
is required to be within a distance of $10$~cm from the detector center.

Tracks with a transverse momentum of $0.16<\pt<1.0$ GeV/$c$ and a pseudorapidity of \mbox{$|\eta|<0.8$} are retained in this analysis. 
To ensure good momentum resolution a minimum of 70 tracking points in the TPC are required.
The measured energy loss (\dEdx) of particles traversing the TPC and the corresponding uncertainty ($\sigma$) are used to select charged pions \cite{Abelev:2014ffa}.
Charged tracks observed in the TPC are identified as pions if their \dEdx\ is within 2$\sigma$ of the Bethe-Bloch expectation for pions while being more than 2$\sigma$ away from the Bethe-Bloch expectation for kaons and protons.
The pion purity in our sample is studied with the HIJING generator \cite{Wang:1991hta}, folded with the ALICE acceptance.
In the sample selected with the procedure described above, about $96\%$ of the particles are expected to be pions.

The effects of track merging and splitting are minimized by rejecting track pairs whose spatial separation in the TPC is smaller than a threshold value \cite{Abelev:2013pqa}.  
For three-pion and four-pion correlations, each same-charge pair in the triplet and quadruplet is required to satisfy this condition.
Oppositely charged pairs are not required to satisfy this cut as they curve in opposite directions in the solenoidal magnetic field and are therefore easily distinguished.  

The low multiplicity events produced in pp and p--Pb collisions contain a non-negligible non-femtoscopic background arising from mini-jets \cite{Aamodt:2010jj,Aamodt:2011kd,Adam:2015pya}.
We reduce this background by retaining only high multiplicity events in pp and p--Pb.
For pp and p--Pb collisions, we retain events with at least 10 and 15 reconstructed charged pions, respectively.
The choice of these boundaries are chosen to provide sufficient statistics while reducing non-femtoscopic background correlations.
The multiplicity cut selects events from the top $46\%$ and $42\%$ of the cross-sections, respectively.
In Pb--Pb collisions, all non-femtoscopic backgrounds are negligible.
We analyze Pb--Pb data from the top $50\%$ collision centrality in ten equally divided intervals.
The collision centrality in Pb--Pb is determined using the charged-particle multiplicity in the V0 detectors~\cite{Abelev:2013qoq}.
Approximately 13, 52, and 34 million events are used for \pp, \pPb, and \PbPb\ collisions, respectively.

\section{Analysis technique}
\label{sec:analysis}
We follow the techniques outlined in Ref.~\cite{Gangadharan:2015ina} for the extraction of multipion QS correlations and a possible coherent fraction.
Several types of multipion correlation functions are presented: $C_3^{\rm QS}$, $c_3^{\rm QS}$, $C_4^{\rm QS}$, $a_4^{\rm QS}$, $b_4^{\rm QS}$, and $c_4^{\rm QS}$.
The full three-pion correlation is given by $C_3^{\rm QS}$ and the cumulant correlation is given by $c_3^{\rm QS}$.
Four types of four-pion correlations are defined: the full correlation, $C_4^{\rm QS}$; two types of partial cumulant correlations, $a_4^{\rm QS}$ and $b_4^{\rm QS}$; and the cumulant correlation, $c_4^{\rm QS}$.

The full three-pion same-charge correlation function contains both pair and triplet symmetrization sequences while the cumulant contains only the triplet symmetrization sequence.
The full four-pion same-charge correlation function contains four sequences of symmetrizations: single-pair, double-pair, triplet, and quadruplet symmetrizations.
Partial cumulants, denoted by $a_4^{\rm QS}$ ($b_4^{\rm QS}$), have single-pair (single- and double-pair) symmetrizations explicitly removed.
The cumulant correlation, denoted by $c_4^{\rm QS}$, represents an isolation of the quadruplet symmetrization sequence.

Two-pion correlations are extracted from two types of pair momentum distributions, $N_1(p_1)N_1(p_2)$ and $N_2(p_1,p_2)$, where $p_i$ is the momentum of particle $i$. 
$N_1(p_1)N_1(p_2)$ is measured by sampling two pions from different events with similar characteristic multiplicity and $z$-coordinate collision vertex class.  
$N_2(p_1,p_2)$ is measured by sampling both pions from the same event.
Three-pion QS correlations are extracted from three types of triplet distributions
\begin{eqnarray}
&N_1(p_1)N_1(p_2)N_1(p_3),& \label{eq:N1N1N1} \\
&N_2(p_1,p_2)N_1(p_3),&  \label{eq:N2N1} \\
&N_3(p_1,p_2,p_3).&  \label{eq:N3}
\end{eqnarray}
Four-pion QS correlations are extracted from the following quadruplet distributions
\begin{eqnarray}
&N_1(p_1)N_1(p_2)N_1(p_3)N_1(p_4),& \label{eq:N1N1N1N1} \\
&N_2(p_1,p_2)N_1(p_3)N_1(p_4),& \label{eq:N2N1N1} \\
&N_2(p_1,p_2)N_2(p_3,p_4),& \label{eq:N2N2} \\
&N_3(p_1,p_2,p_3)N_1(p_4),& \label{eq:N3N1} \\
&N_4(p_1,p_2,p_3,p_4).& \label{eq:N4}
\end{eqnarray}
The distributions in Eqs.~\ref{eq:N1N1N1}-\ref{eq:N4} are formed by sampling the appropriate number of particles from the same event and the rest from different events.
The subscript for $N$ represents the number of pions taken from the same event.
We normalize the distributions in Eqs.~\ref{eq:N1N1N1}-\ref{eq:N2N1} to the distribution in Eq.~\ref{eq:N3} at a suitably large invariant relative momentum, $q_{ij} = \sqrt{-(p_i-p_j)^{\mu}(p_i-p_j)_{\mu}}$.
Likewise, the distributions in Eqs.~\ref{eq:N1N1N1N1}-\ref{eq:N3N1} are normalized to the distribution in Eq.~\ref{eq:N4}.
The $q_{ij}$ interval is chosen to be far away from the region of significant QS and FSI correlations.  
The normalization interval is $0.15<q_{ij}<0.2$ GeV/$c$ in Pb--Pb while being $0.9<q_{ij}<1.2$ GeV/$c$ in pp and p--Pb due to the wider QS correlations in smaller collision systems.
The distributions are all corrected for finite momentum resolution and muon contamination \cite{Abelev:2013pqa}.

The two-, three-, and four-pion distributions ($N_n^{\rm QS}$) are extracted from the measured distributions ($N_n$) with the appropriate coefficients according to the ``core-halo'' prescription \cite{Csorgo:1994in} of short- and long-lived emitters \cite{Lednicky:1979ig}.  
In the core-halo model, a fraction of particles ($f_c$) originate within a small radius component of particle production (the core).  
The rest, $1-f_c$, originate within a much larger halo radius.
The fraction of pairs, triplets, and quadruplets from the core is then given by $f_c^2$, $f_c^3$, and $f_c^4$, respectively.
The other possibilities of mixed core-halo compositions are treated as well in this analysis.
Pairs of particles from the core of particle production are separated by sufficiently short distances such that their QS and FSI correlations are experimentally observable.  
Pairs with one or both particles from the halo effectively dilute the correlation functions as no significant QS and FSI correlations are expected.  
The coefficients that isolate the multipion QS distributions are determined from the $f_c$ parameter \cite{Gangadharan:2015ina}.

The $f_c$ parameter is often associated with $\sqrt{\lambda}$, where $\lambda$ parametrizes the correlation strength, which is usually determined from fits to two-particle Bose-Einstein correlations.
However, due to the unknown functional form of two-pion correlation functions, the $\lambda$ parameter, determined this way, is convoluted with the arbitrary choice of fitting functions (e.g. Gaussian fits to non-Gaussian correlation functions).
A more accurate extraction of $f_c$ is done by fitting mixed-charge two-pion correlations instead \cite{Abelev:2013pqa}.  
The correlation between $\pi^+$ and $\pi^-$ is dominated by Coulomb and strong FSI for which the wave functions are well known \cite{Lednicky:2005tb}.
Owing to the large pion Bohr radius, $\pi^+\pi^-$ correlations are less sensitive to the detailed structure of the source and can be fit less ambiguously wrt $\pi^+\pi^+$ correlations.
As part of the long-lived emitters correspond to weak decays (secondaries), $f_c$ is also sensitive to the specific tracking algorithm's ability to discriminate primary from secondary tracks. 
The value, $f_c=0.84\pm0.03$, was used in Ref.~\cite{Abelev:2013pqa} as well as in this analysis. 

The distinction between core and halo may depend on the characteristic sizes and the dynamics of the system.  Pions from decays of mid-lived emitters, such as the  $K^*$, $\Sigma^*$, $\omega$, and $\eta'$ constitute a special case where the effect of QS correlations with other pions can be smaller than that of Coulomb correlations.  Therefore, one might expect a slightly smaller core fraction for QS compared to Coulomb interactions.  The magnitude of the difference should mainly relate to the fraction of pions produced from decays of mid-lived resonances.  The resulting difference, which we assume to be small, is addressed by varying $f_c$ as discussed in the section on systematic uncertainties.

The treatment of multibody FSI (Coulomb and strong) is done according to the generalized Riverside approximation \cite{Liu:1986nb,Aggarwal:2000ex,Adams:2003vd,Abelev:2013pqa,Gangadharan:2015ina} where the $n$ body FSI correlation is treated as the product of each pair FSI correlation,
\begin{eqnarray}
&K_3 = K_2(q_{12})K_2(q_{13})K_2(q_{23}),& \\
&K_4 = K_2(q_{12})K_2(q_{13})K_2(q_{14})K_2(q_{23})K_2(q_{24})K_2(q_{34}).&
\end{eqnarray}
The two-pion FSI factor of pair $(i,j)$ is given by $K_2(q_{ij})$ and is calculated by averaging the modulus square of the Coulomb and strong wave function over an assumed freeze-out distribution.
We use the \textsc{therminator} model of particle production as an estimate for the freeze-out distribution \cite{Kisiel:2005hn,Chojnacki:2011hb}.
The pair product approach to three-pion FSI correlations was shown to be a good approximation to the full asymptotic wave function calculation \cite{Abelev:2013pqa,Gangadharan:2015ina}.
In this article we present QS correlation functions which are corrected for FSI and for the dilution of long-lived emitters according to Eqs.~33 and 39 in Ref.~\cite{Gangadharan:2015ina}.

All distributions and correlation functions are projected onto the 1D the Lorentz invariant relative momentum.
For three- and four-pion correlations, the sum quadrature of pair invariant relative momenta is used:
\begin{eqnarray}
Q_3 &=& \sqrt{q_{12}^2+q_{13}^2+q_{23}^2}, \\
Q_4 &=& \sqrt{q_{12}^2+q_{13}^2+q_{14}^2+q_{23}^2+q_{24}^2+q_{34}^2}.
\end{eqnarray}
The \pt~dependence of the correlation functions is studied by further projecting onto the average transverse momenta
\begin{eqnarray}
\kT &=& \frac{|\vec{p}_{\rm T,1}+\vec{p}_{\rm T,2}|}{2}, \\
\KTThree &=& \frac{|\vec{p}_{\rm T,1}+\vec{p}_{\rm T,2}+\vec{p}_{\rm T,3}|}{3}, \\
\KTFour &=& \frac{|\vec{p}_{\rm T,1}+\vec{p}_{\rm T,2}+\vec{p}_{\rm T,3}+\vec{p}_{\rm T,4}|}{4},
\end{eqnarray}
for two-, three-, and four-pion correlations, respectively.
We form two intervals of \KTThree~defined by $0.16<\KTThree<0.3$ and $0.3<\KTThree<1.0$ GeV/$c$.
Similarly, we define two intervals of \KTFour~as $0.16<\KTFour<0.3$ and $0.3<\KTFour<1.0$ GeV/$c$.
For the low \KTThree~interval which is simultaneously at low $Q_3$ ($0.02<Q_3<0.03$ GeV/$c$), $\left< \pt \right>=0.23$ GeV/$c$ and the RMS of the \pt~distribution is 0.03 GeV/$c$.
At high \KTThree, $\left< \pt \right>$ is 0.34 GeV/$c$ and the RMS is 0.03 GeV/$c$.
The same values also closely describe the low and high \KTFour~interval at low $Q_4$ ($0.045<Q_4<0.06$ GeV/$c$).
We further note that the $\left< \pt \right>$ is very similar for each \qinv~interval in this analysis.  
For $0.16<\kT<0.3$ GeV/$c$, $\left< \pt \right>$ increases linearly by about 0.015 GeV/$c$ in the interval $0.005<\qinv<0.2$ GeV/$c$.

\begin{figure}[t!p]
  \centering
  \subfigure[pp low \KTFour]{
    \includegraphics[width=0.48\textwidth]{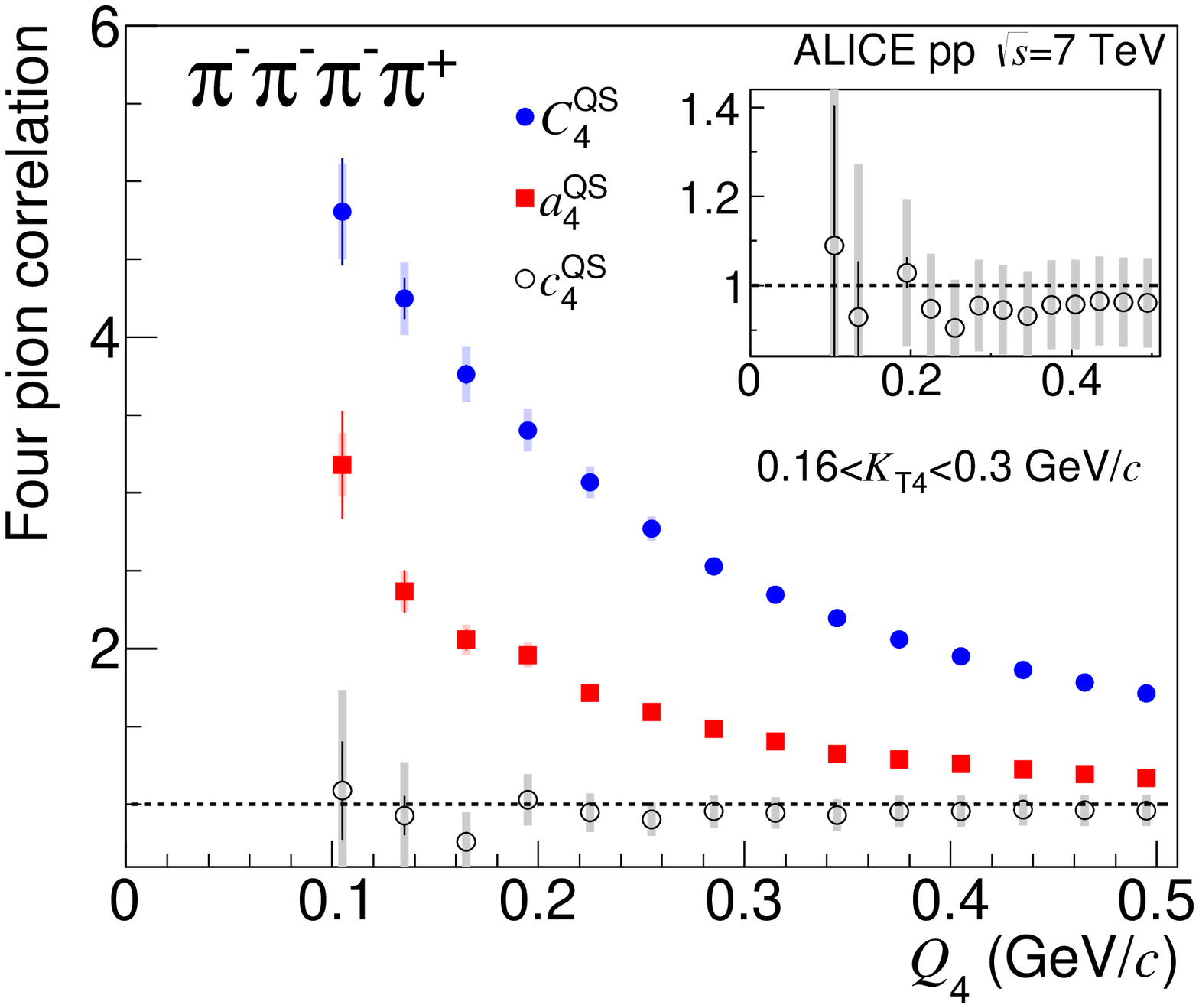}
    \label{fig:C4MC1_pp_K0}
  }
  \subfigure[pp high \KTFour]{
    \includegraphics[width=0.48\textwidth]{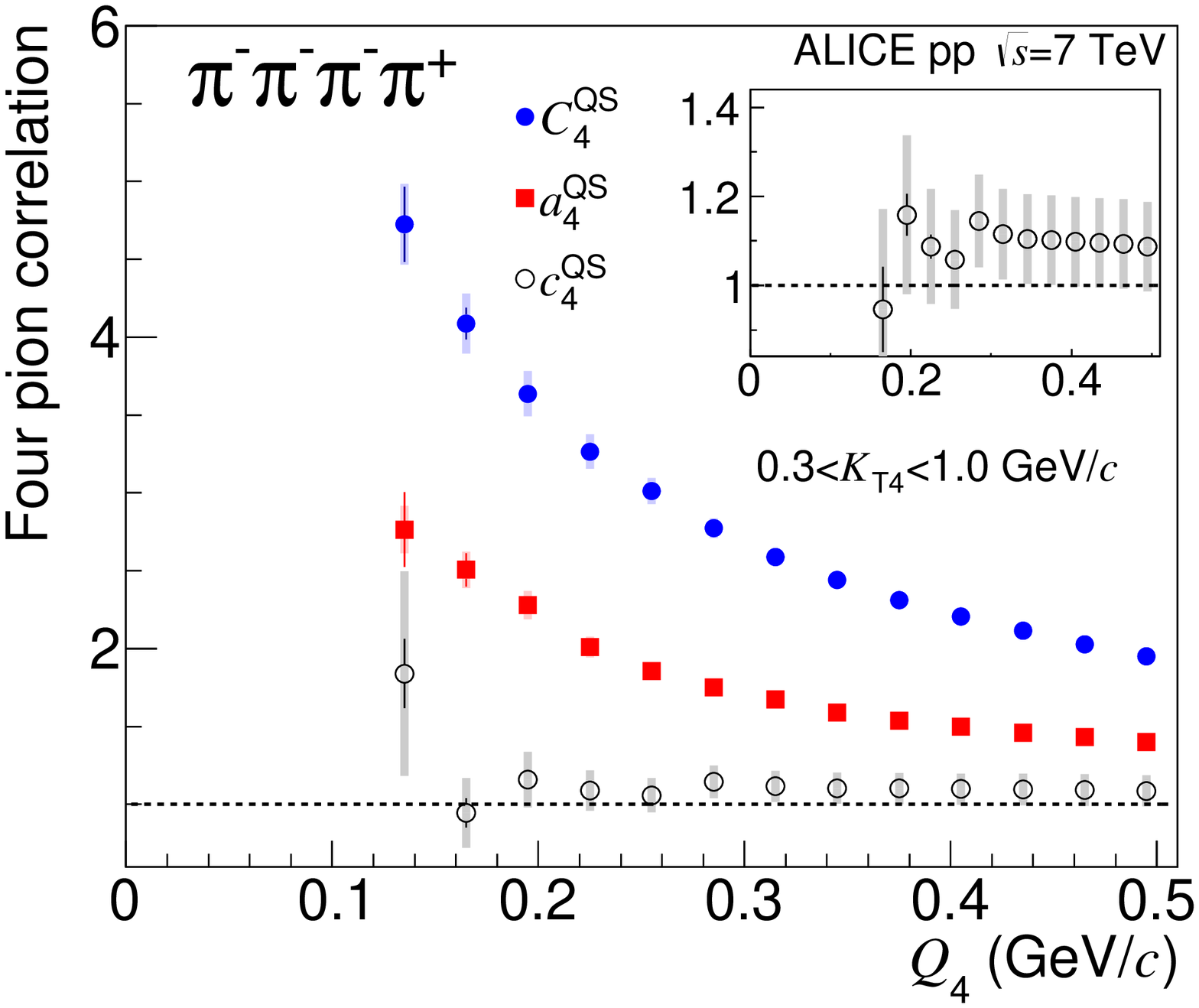}
    \label{fig:C4MC1_pp_K1}
  }

  \subfigure[p--Pb low \KTFour]{
    \includegraphics[width=0.48\textwidth]{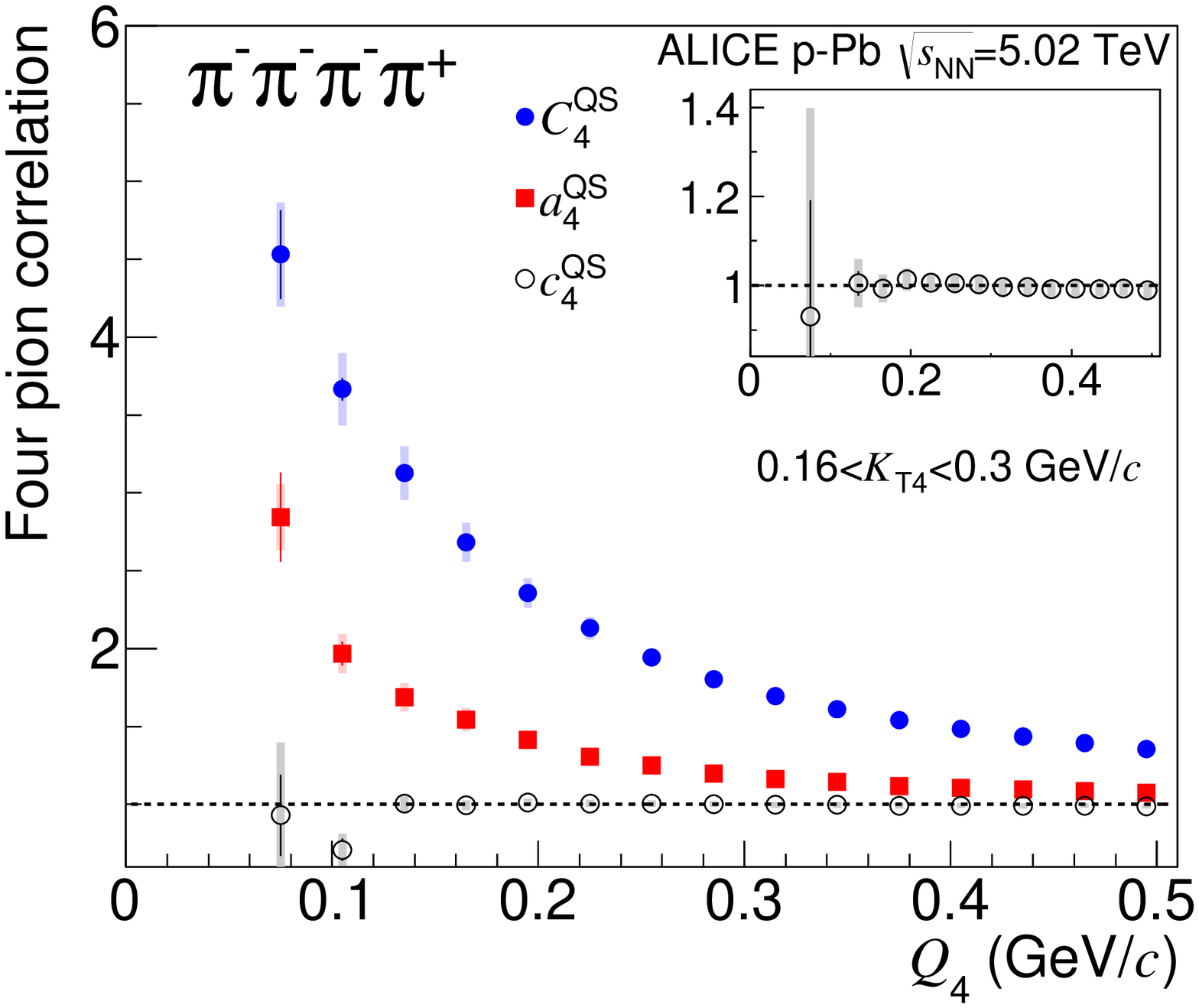}
    \label{fig:C4MC1_pPb_K0}
  }
  \subfigure[p--Pb high \KTFour]{
    \includegraphics[width=0.48\textwidth]{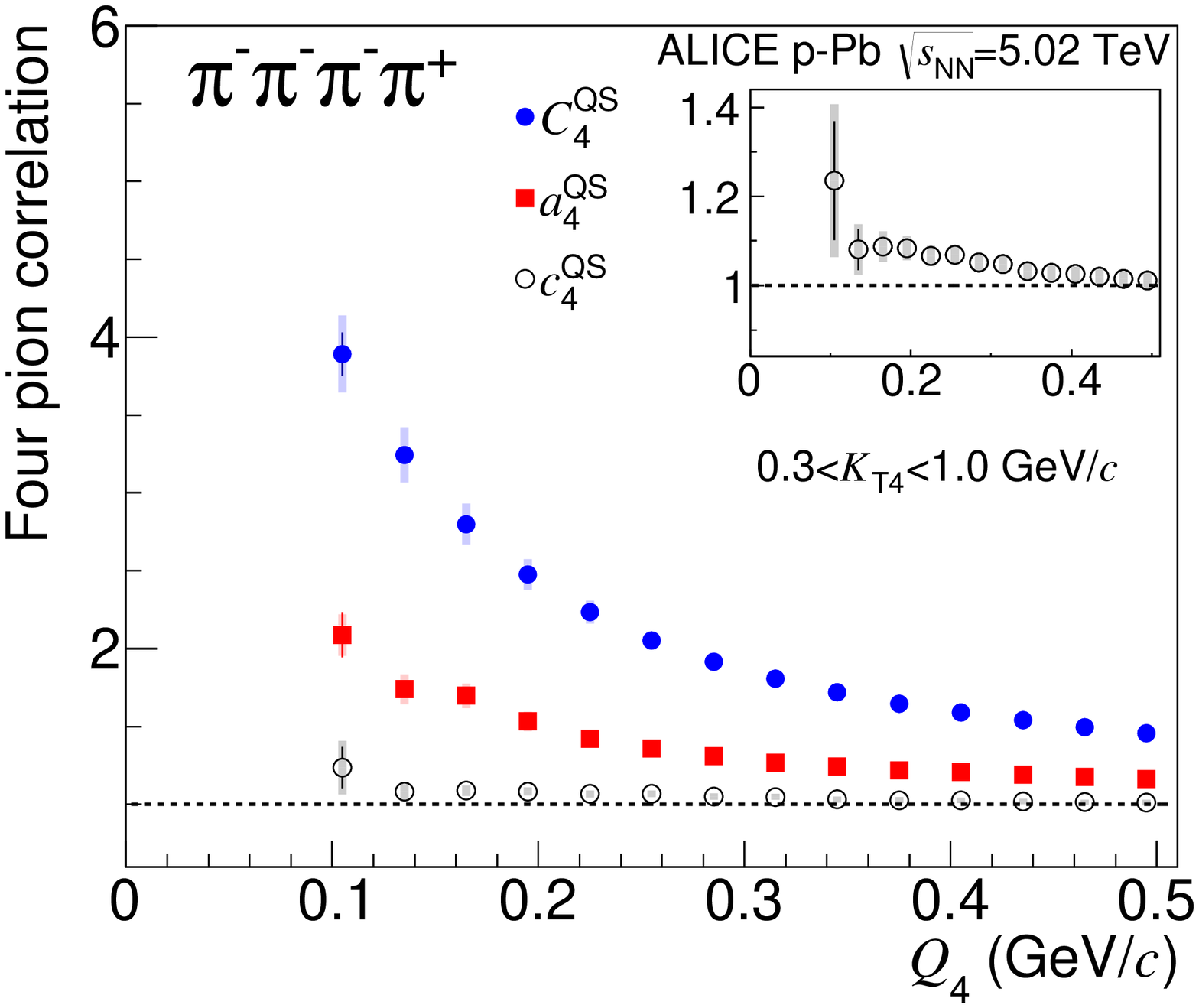}
    \label{fig:C4MC1_pPb_K1}
  }

  \subfigure[Pb--Pb low \KTFour]{
    \includegraphics[width=0.48\textwidth]{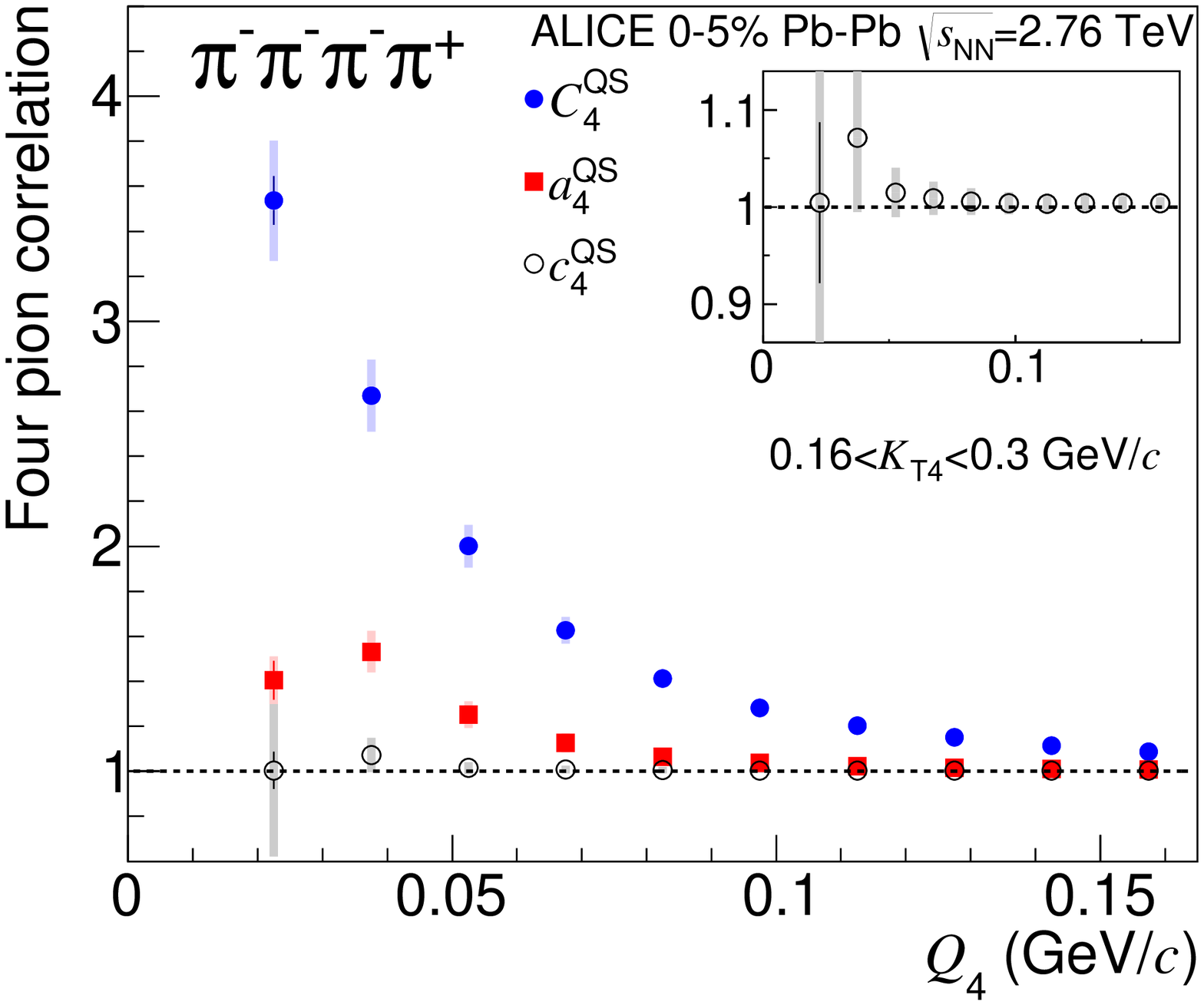}
    \label{fig:C4MC1_PbPb_K0}
  }
  \subfigure[Pb--Pb high \KTFour]{
    \includegraphics[width=0.48\textwidth]{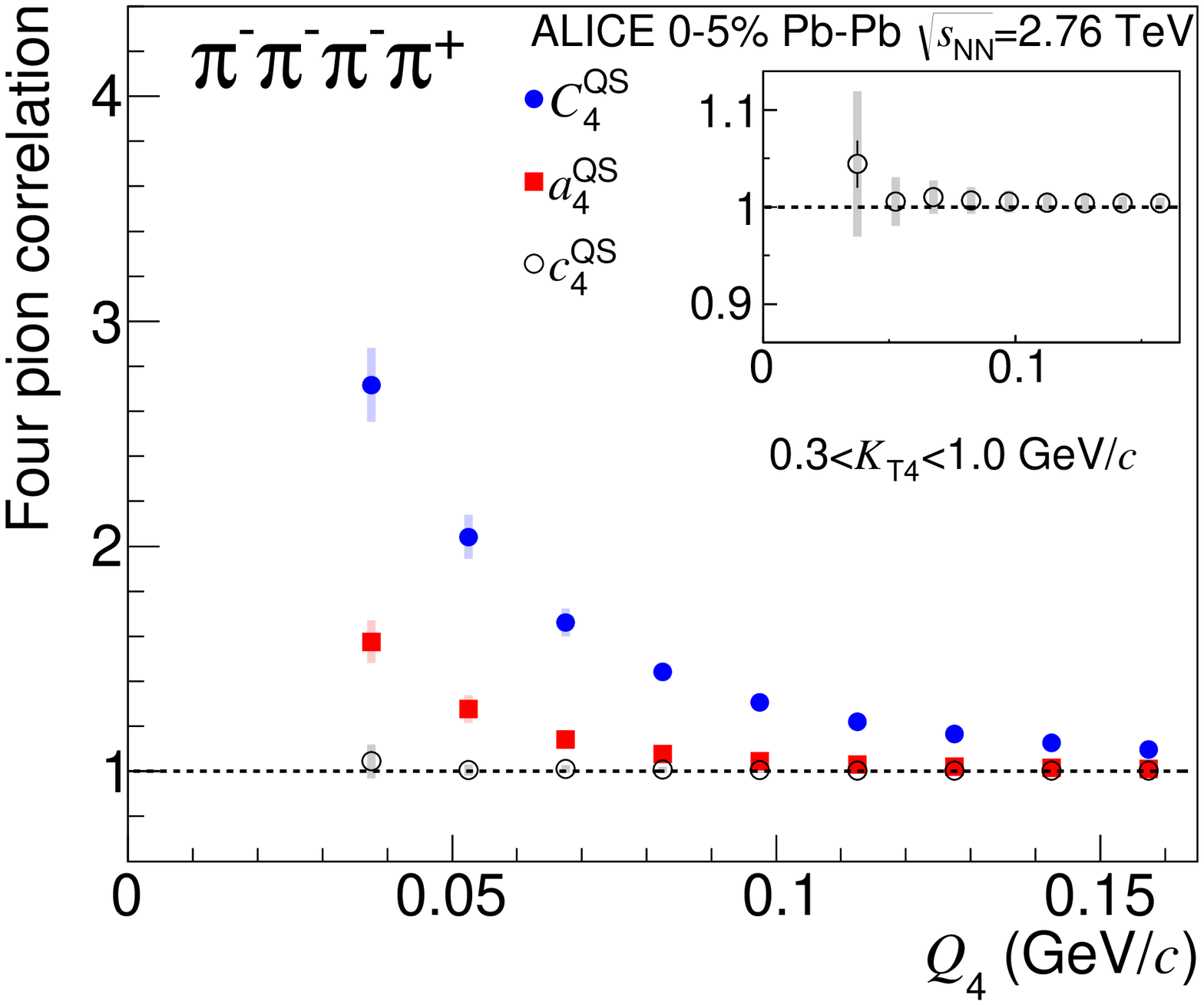}
    \label{fig:C4MC1_PbPb_K1}
  }
  \caption{Mixed-charge ($\pm\pm\pm\mp$) four-pion correlations versus $Q_4$ in pp, p--Pb, and Pb--Pb.  The full ($C_4^{\rm QS}$), partial cumulant ($a_4^{\rm QS}$), and cumulant ($c_4^{\rm QS}$) correlation functions are shown.  The inset figure shows a zoom of $c_4^{\rm QS}$. Systematic uncertainties are shown by the shaded bands.  Low and high \KTFour~quadruplets are shown.  The average of the charge conjugated correlation functions is shown.}
\end{figure}

\subsection{Extracting the pair-exchange magnitudes}
The building blocks of Bose-Einstein correlations are the pair-exchange magnitudes ($T_{ij}$) and the coherent fraction ($G$) in the absence of multipion phases \cite{Andreev:1992pu,Heinz:1997mr,Csorgo:1999sj,Gangadharan:2015ina}.
Multipion phases are expected when the space-time point of maximum pion emission is momentum dependent.  
However, the relative momentum dependence of the effect was shown to be rather weak \cite{Heinz:1997mr}.
Assuming a value of $G$, the pair-exchange magnitudes can be used to build all higher orders of correlation functions.
We define the {\it expected} or {\it built} correlation functions, $E_n(i)$, which represent the expectation of higher order ($n$) QS correlations using lower order ($i<n$) experimental measurements as an input.
The equations to build $E_n$ are given in appendix \ref{app:supp}.
We define two types of expected correlation functions:
\begin{enumerate}
\item \underline{$E_3(2)$ and $E_4(2)$}: The pair-exchange magnitudes can be extracted directly from two-pion correlation functions, which forms our primary expectation in Pb--Pb collisions.  
The two-pion correlations are tabulated in four dimensions during the first pass over the data in the longitudinally co-moving system ($q_{\rm out}, q_{\rm side}, q_{\rm long}, \kT$). 
The interval width of each relative momentum dimension is 5 MeV/$c$, while it is 50 MeV/$c$ in the \kT~dimension.
In the second pass over the data, the previously tabulated two-pion correlations are interpolated for each pion pair from mixed events.
We interpolate between relative momentum bins with a cubic interpolator.  
A linear interpolation is used in between \kT~bins, where a more linear dependence of correlation strength is observed.
\item \underline{$E_4(3)$ and $e_4(3)$}: We also extract the pair-exchange magnitudes from fits to $C_3^{\rm QS}$ ($E_4(3)$) and $c_3^{\rm QS}$ ($e_4(3)$) in 3D ($q_{12},q_{13},q_{23}$). 
The fit is performed according to an Edgeworth parametrization \cite{Csorgo:2000pf} as shown in equation 20 of Ref.~\cite{Gangadharan:2015ina}. 
This $2^{\rm nd}$ approach is more limited as the pair-exchange magnitudes are extracted from a 3D projection of a 9D function. 
Similar to the $1^{\rm st}$ type of expected correlations, the pair-exchange magnitudes are obtained from the first pass over the data and input into the second pass.
\end{enumerate}

For the case of partial coherence, we assume that the pair-exchange magnitude of the coherent source is identical to the chaotic one (e.g.\ same radii) which might be expected for DCC radiation \cite{Akkelin:2001nd}.
The value of $G$ may then be extracted by minimizing the $\chi^2$ difference between measured and expected correlations for each $Q_3$ or $Q_4$ bin.
One may extract $G$ from either of the six same-charge channels: $C_4^{\rm QS}$, $a_4^{\rm QS}$, $b_4^{\rm QS}$, $c_4^{\rm QS}$, $C_3^{\rm QS}$, and $c_3^{\rm QS}$.  
The primary channel of extraction is $C_4^{\rm QS}$ for reasons of statistical precision and sensitivity to coherent emission.  
We also extracted $G$ with several other multipion correlations and is shown in a separate note \cite{pubnote:2015}.
In pp and p--Pb collisions, where non-negligible non-femtoscopic backgrounds exist, we only use the $2^{\rm nd}$ build technique as three-pion correlations have a larger signal to background ratio \cite{Abelev:2014pja}.

Both build techniques were tested using data generated by the \textsc{therminator} model, including a known coherent fraction \cite{Gangadharan:2015ina}.  
The $E_4(2)$ correlations were typically $3\%$ smaller than the ``measured correlations'' in \textsc{therminator}.  
The bias is attributed to the finite 4D projection of the true 6D two-pion correlation function.
We correct for this potential bias in a data-driven approach.  
The interpolated two-pion correlation function from the 4D projection is compared to the true two-pion correlation function for each $\qinv$ interval.  
The ratio of the two correlation functions (subtracting unity from each), forms our correction factor.

\section{Results}
\label{sec:results}
We now present the results of three- and four-pion QS correlations in pp, p--Pb, and Pb--Pb collisions.
All correlations are corrected for FSI and for the dilution of pions from long-lived emitters.  
Mixed-charge correlations are first presented to demonstrate the effectiveness of all corrections in the analysis.
Fits to same-charge three-pion correlations, which allow us to construct $E_4(3)$ and $e_4(3)$, are then presented.  
The comparison of measured to expected same-charge correlations assuming the null hypothesis ($G=0$) is then presented.
Finally we present the same comparison with non-zero values of $G$.

\subsection{Mixed-charge four-pion correlation functions}
Mixed-charge correlations of the first type ($\pm\pm\pm\mp$) are shown in Figs.~\ref{fig:C4MC1_pp_K0}-\ref{fig:C4MC1_PbPb_K1}.
The full correlation contains contributions from two- and three-pion symmetrizations while the partial cumulant ($a_4^{\rm QS}$) contains only three-pion symmetrizations.
The cumulant ($c_4^{\rm QS}$) has all lower orders ($n<4$) of symmetrization removed.
Its proximity to unity demonstrates the effectiveness of several procedures: the event-mixing technique, FSI corrections, muon corrections, and momentum resolution corrections.

\begin{figure}[t!h]
  \centering
  \subfigure[pp low \KTFour]{
    \includegraphics[width=0.48\textwidth]{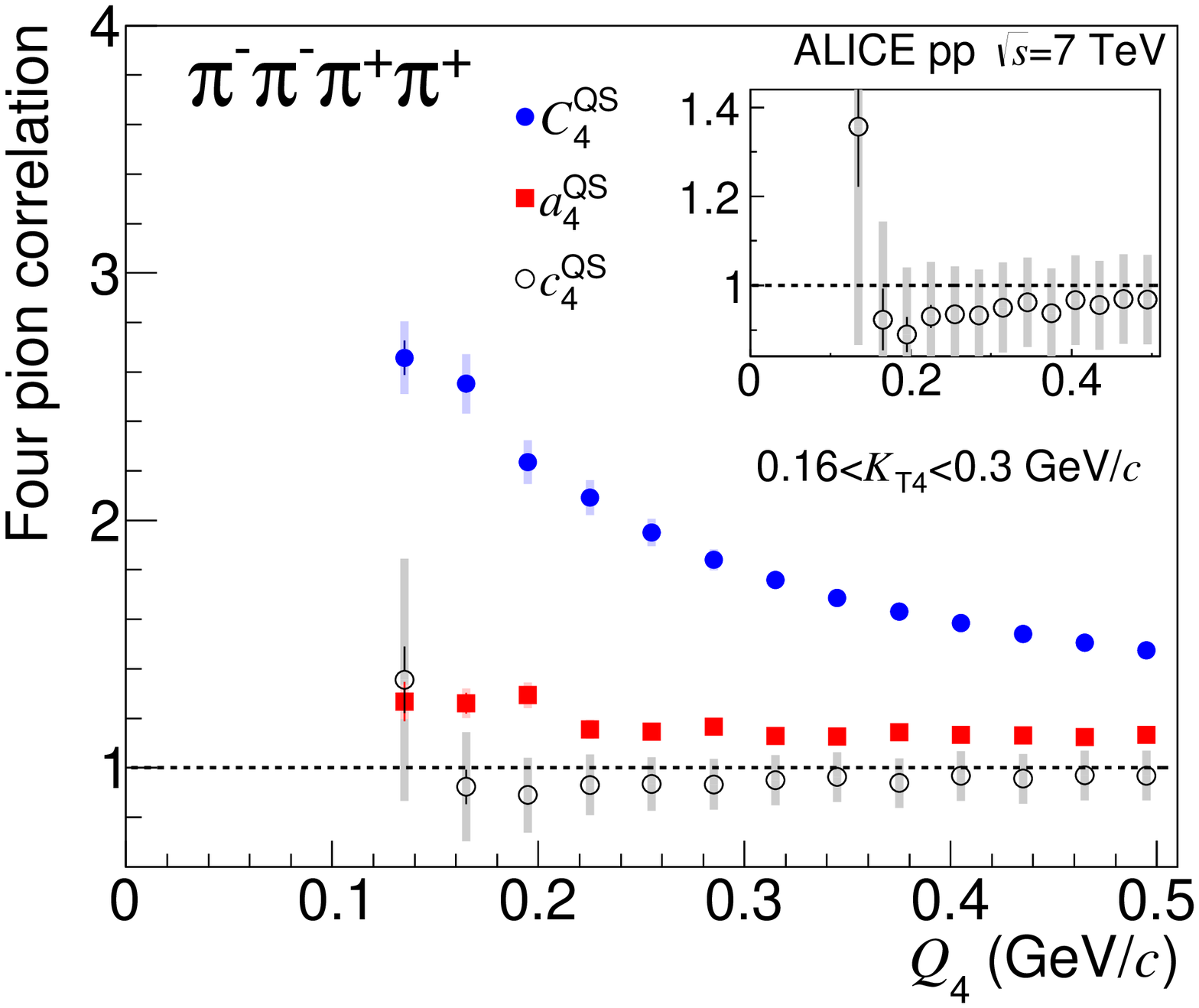}
    \label{fig:C4MC2_pp_K0}
  }
  \subfigure[pp high \KTFour]{
    \includegraphics[width=0.48\textwidth]{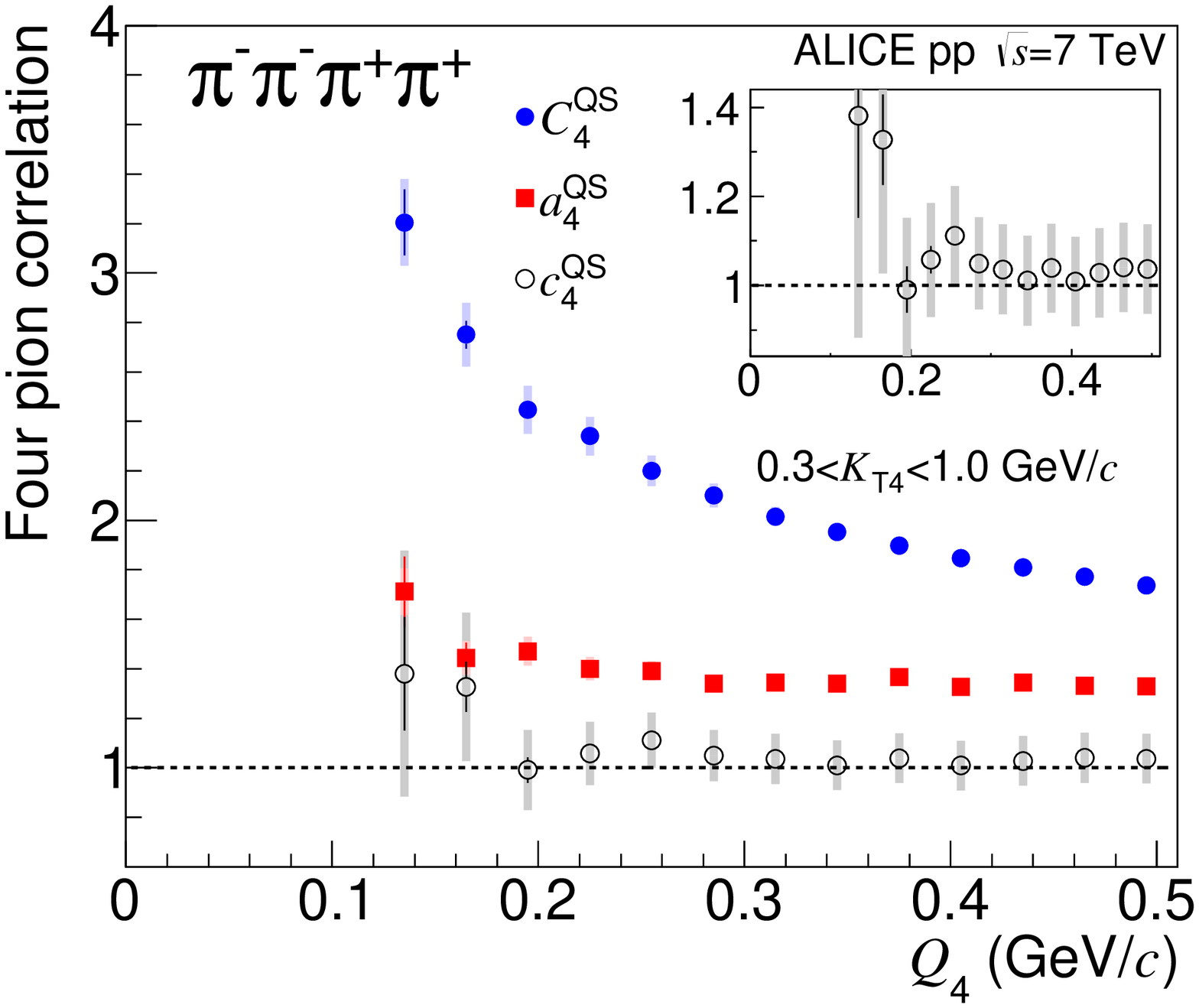}
    \label{fig:C4MC2_pp_K1}
  }

  \subfigure[p--Pb low \KTFour]{
    \includegraphics[width=0.48\textwidth]{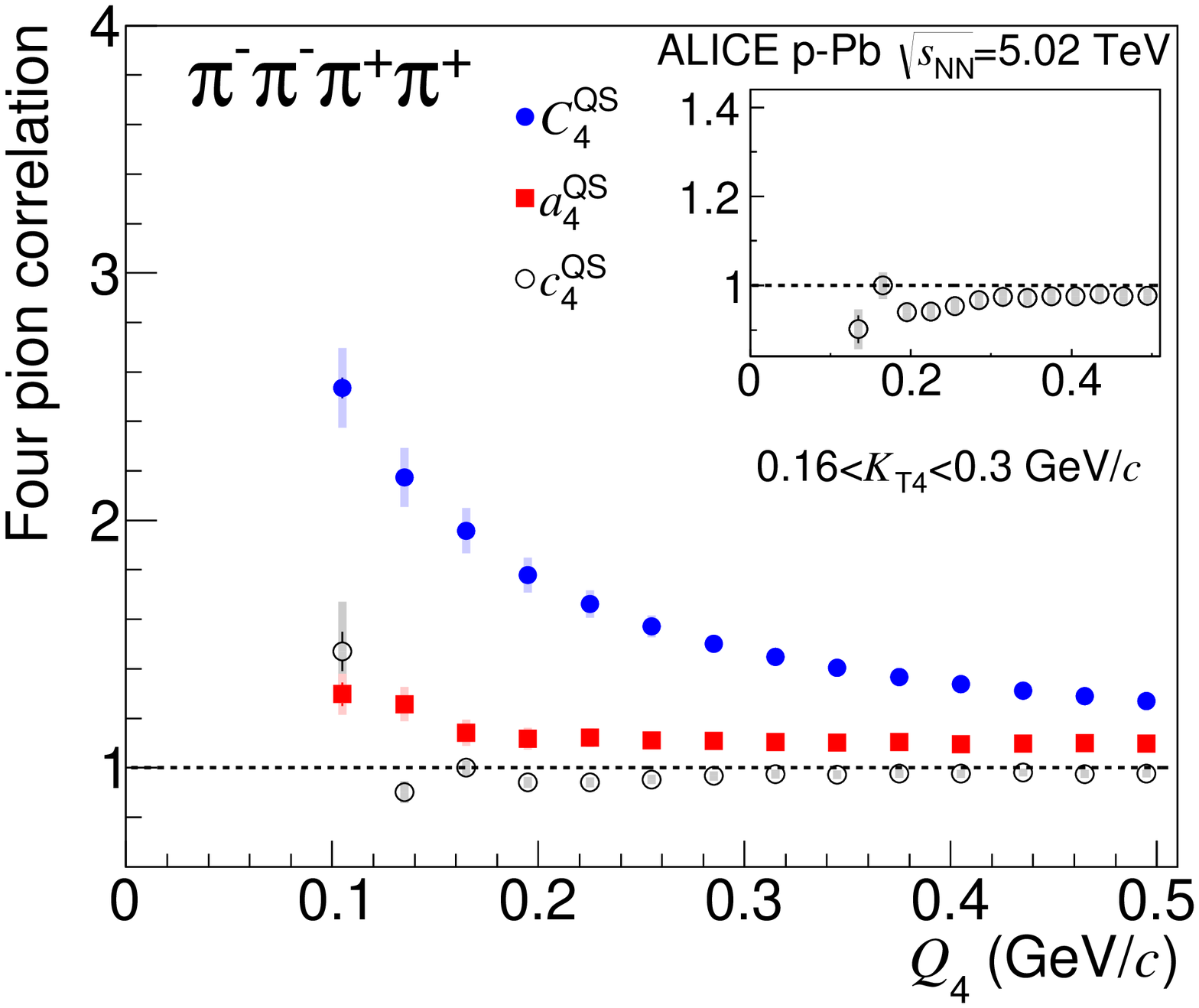}
    \label{fig:C4MC2_pPb_K0}
  }
  \subfigure[p--Pb high \KTFour]{
    \includegraphics[width=0.48\textwidth]{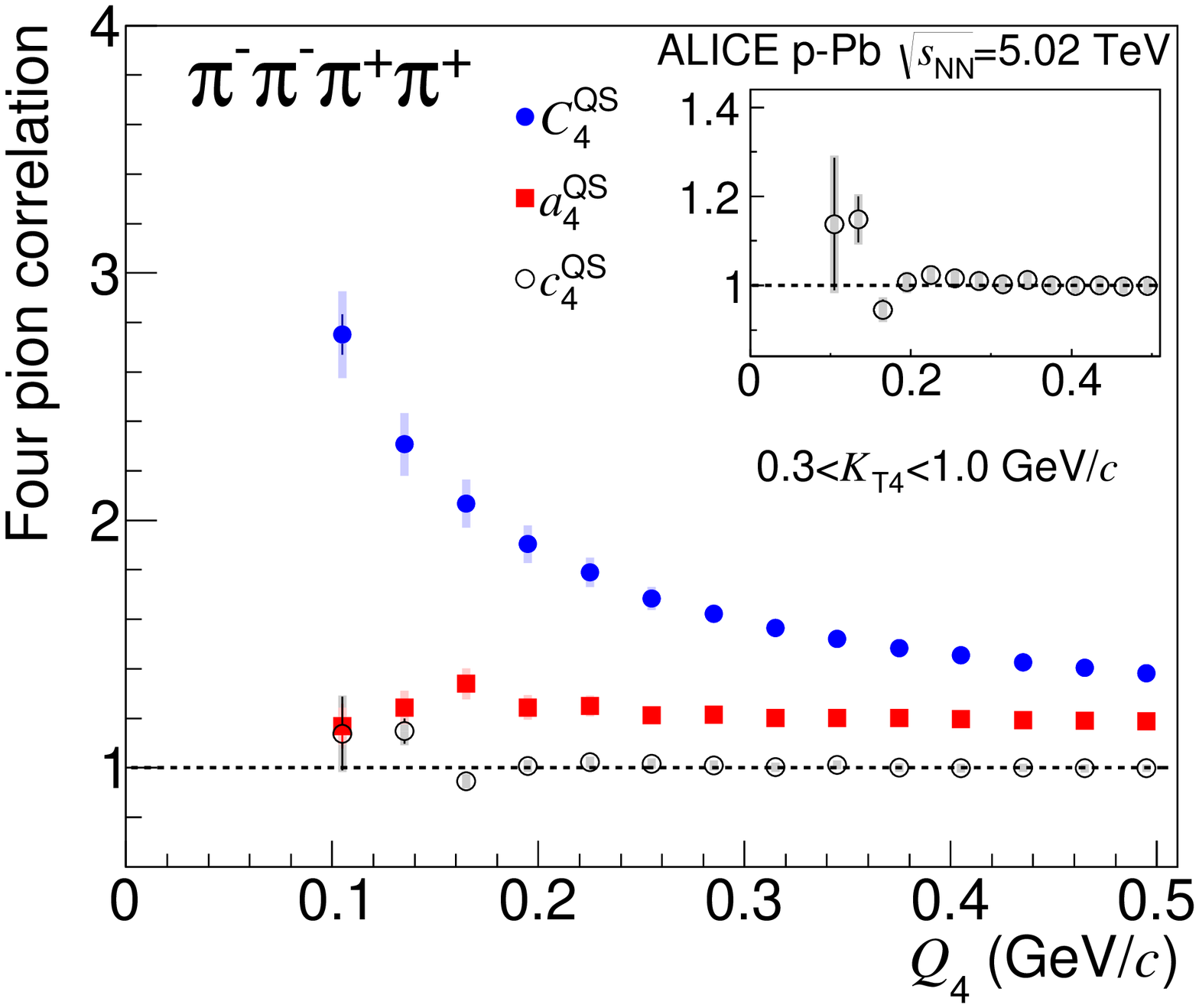}
    \label{fig:C4MC2_pPb_K1}
  }

  \subfigure[Pb--Pb low \KTFour]{
    \includegraphics[width=0.48\textwidth]{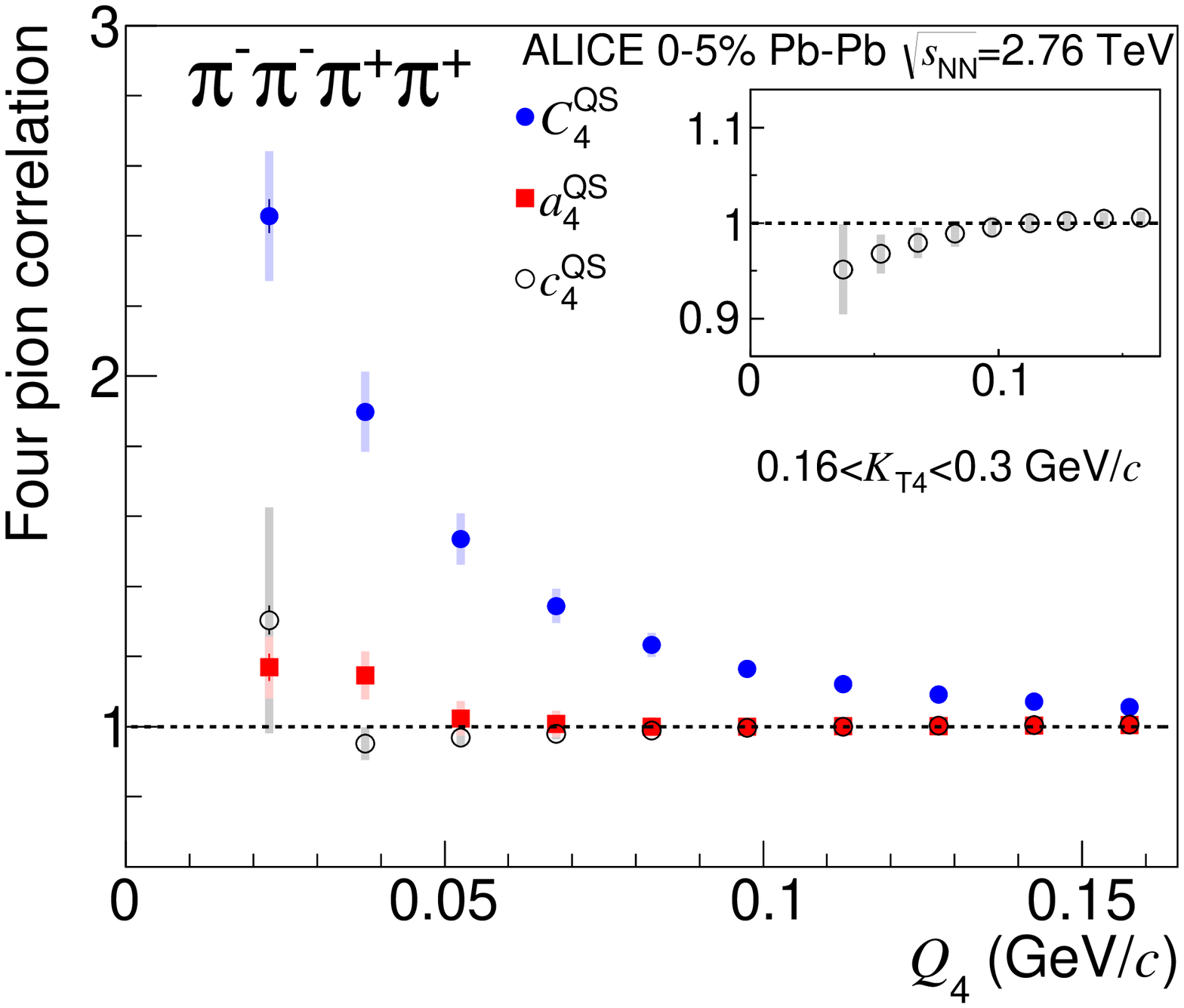}
    \label{fig:C4MC2_PbPb_K0}
  }
  \subfigure[Pb--Pb high \KTFour]{
    \includegraphics[width=0.48\textwidth]{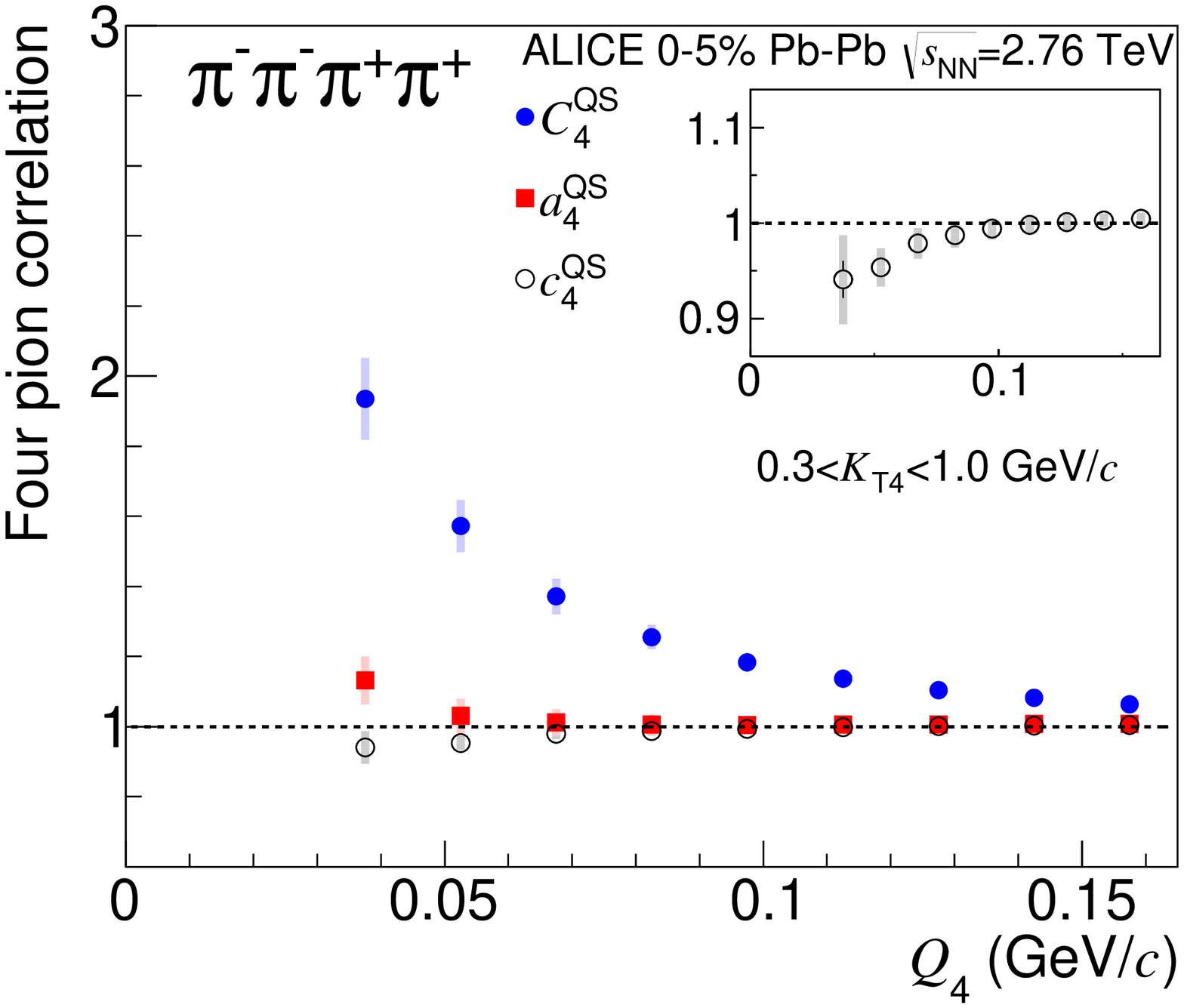}
    \label{fig:C4MC2_PbPb_K1}
  }
  \caption{Mixed-charge ($\mp\mp\pm\pm$) four-pion correlations versus $Q_4$ in pp, p--Pb, and Pb--Pb.  Same details as for Figs.~\ref{fig:C4MC1_pp_K0}-\ref{fig:C4MC1_PbPb_K1}.}
\end{figure}

\begin{figure}[t!h]
  \centering
  \subfigure[pp]{
    \includegraphics[width=0.48\textwidth]{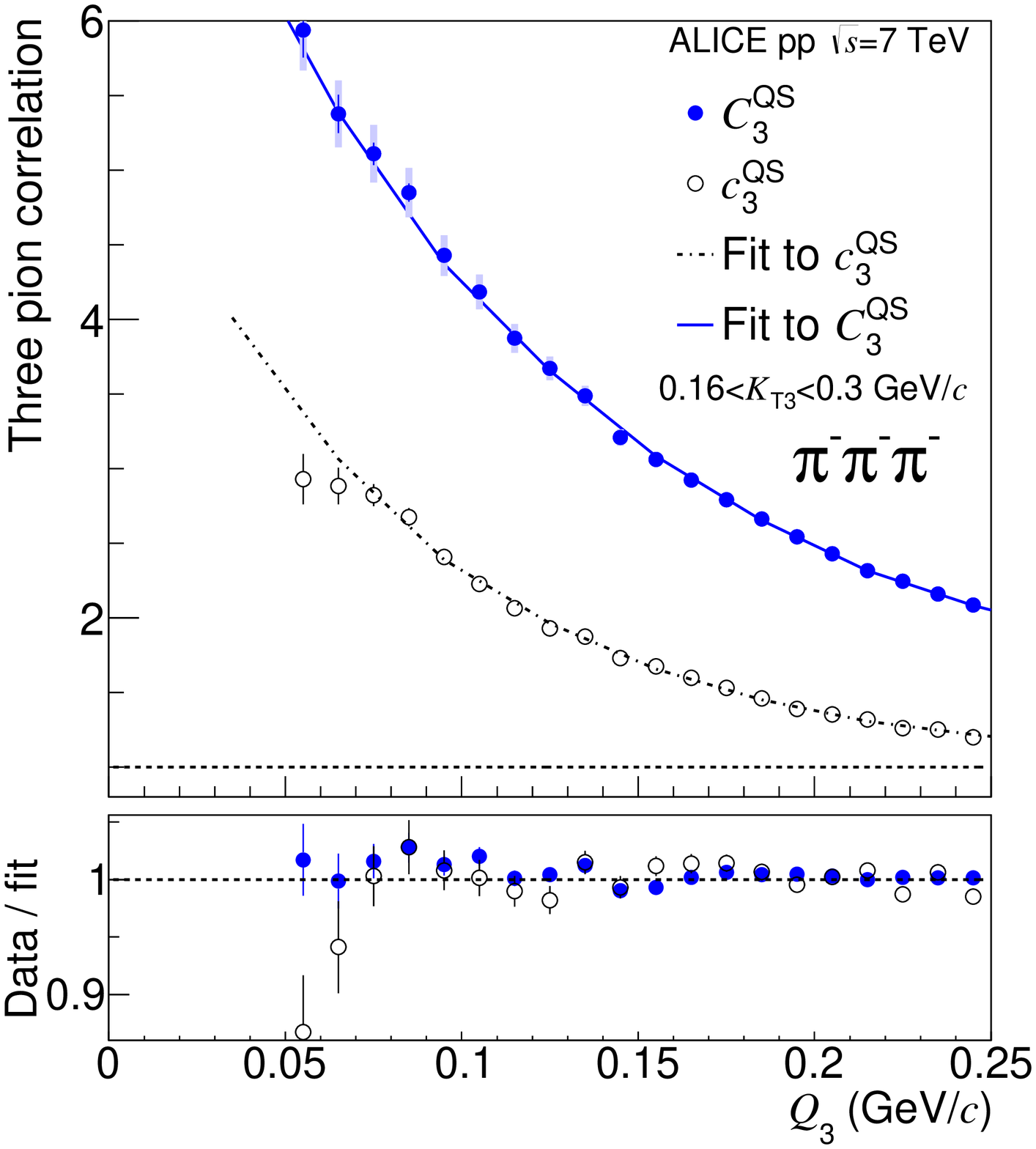}
    \label{fig:C3fits_pp_K0}
  }
  \subfigure[p--Pb]{
    \includegraphics[width=0.48\textwidth]{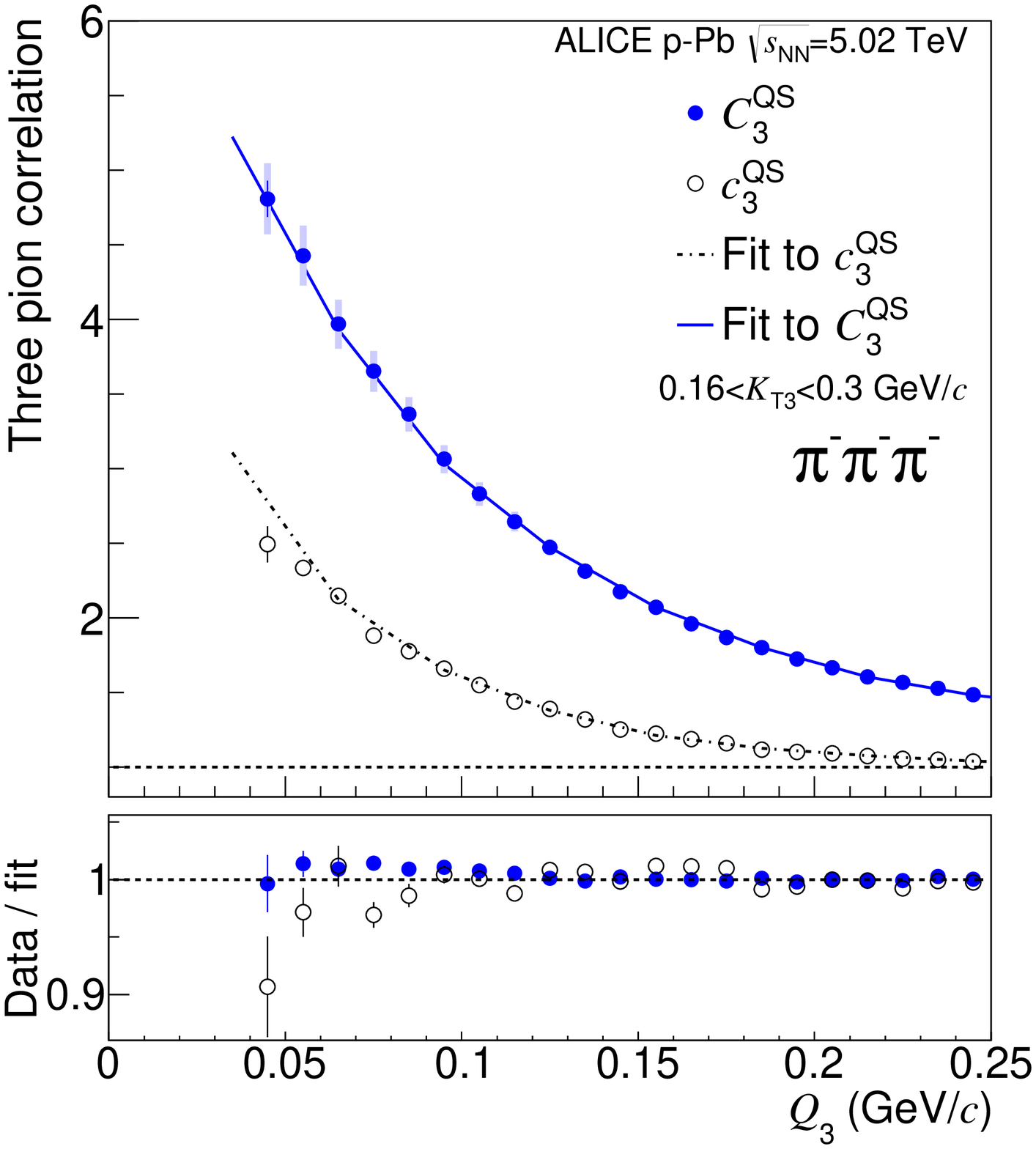}
    \label{fig:C3fits_pPb_K0}
  }
  \subfigure[Pb--Pb]{
    \includegraphics[width=0.48\textwidth]{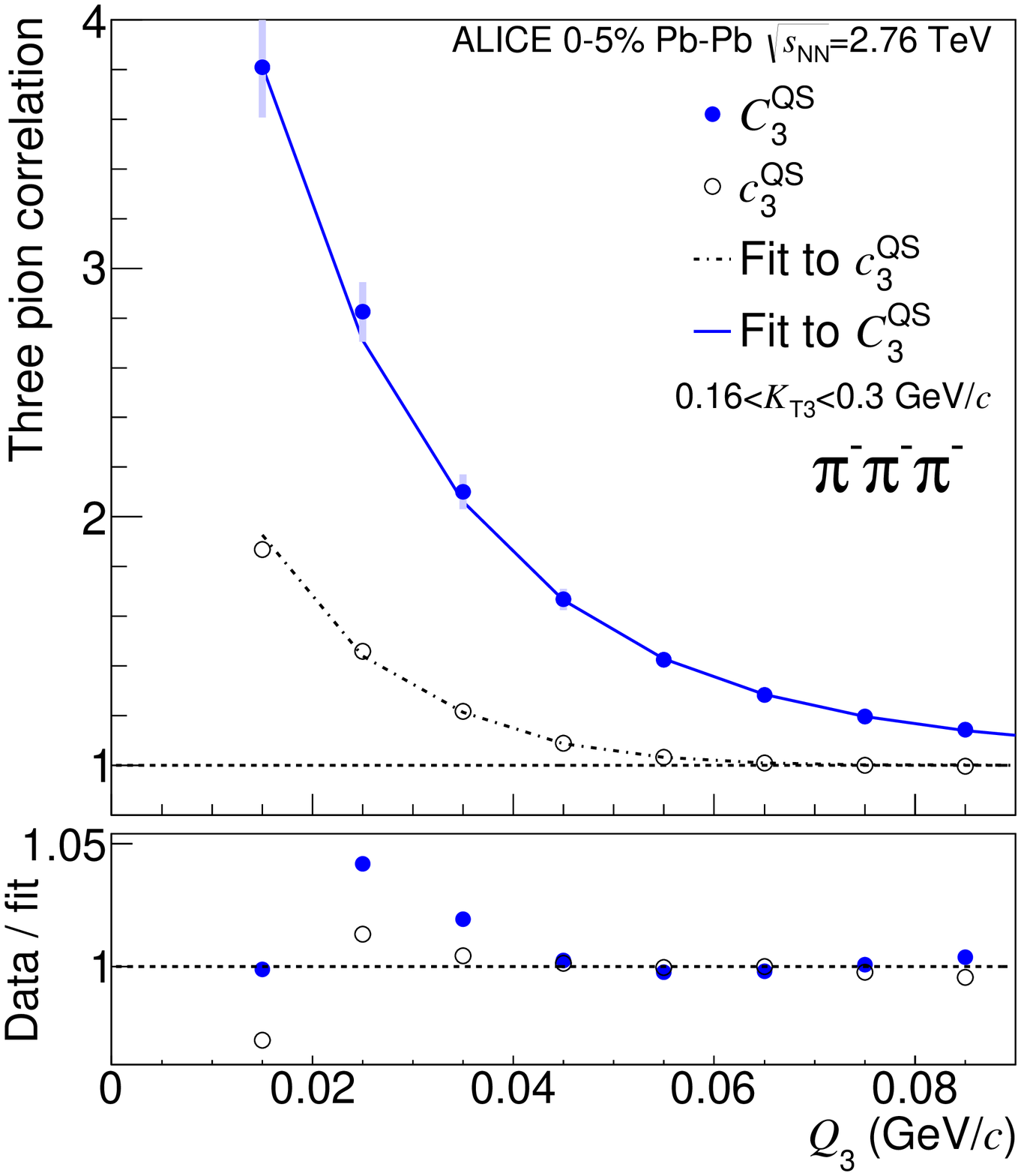}
    \label{fig:C3fits_PbPb_K0}
  }
  \caption{Same-charge three-pion full and cumulant correlations versus $Q_3$ with Edgeworth fits in pp, p--Pb, and Pb--Pb collisions.  Bottom panels show the ratio of the data to the fit.  The fits assume $G=0$.  The systematic uncertainties for $C_3^{\rm QS}$ are given by the shaded band while those for $c_3^{\rm QS}$ are the same after re-scaling by the ratio of correlation strengths.  Only statistical errors are shown for the ratio. The average of the charge conjugated correlation functions is shown.}
\end{figure}

The second type of mixed-charge quadruplets ($\mp\mp\pm\pm$) are shown in Figs.~\ref{fig:C4MC2_pp_K0}-\ref{fig:C4MC2_PbPb_K1}.
The full correlation in Figs.~\ref{fig:C4MC2_pp_K0}-\ref{fig:C4MC2_PbPb_K1} contains contributions from single-pair and double-pair symmetrization sequences.  
The partial cumulant removes the two-pion symmetrizations while the cumulant further removes the double-pair symmetrizations.
Just as for the first type of mixed-charge quadruplets, the residue seen with the cumulant characterizes the effectiveness of several procedures.
The baseline of the cumulant in pp collisions is offset from unity by about 10$\%$ and is due to statistical fluctuations in the high $q$ normalization region of our data sample.
It is included in the systematic uncertainty.
The mixed-charge cumulant residues seen in pp and p--Pb collisions are similar in magnitude as seen in Pb--Pb collisions.
Note that the FSI correlations are larger in pp and p--Pb with respect to Pb--Pb collisions.
Isolation of the cumulant correlation function, $c_4^{\rm QS}$, is done by subtracting several distributions as shown in Eqs.~\ref{eq:N1N1N1N1}-\ref{eq:N3N1} after correcting for FSI.
By default, we also utilize the distributions of two interacting opposite charge pions, $N_2(-,+)N_1(-)N_1(-)$ and $N_2(-,+)N_1(-)N_1(+)$ for $\pi^- \pi^- \pi^- \pi^+$ and  $\pi^- \pi^- \pi^+ \pi^+$, respectively.
After correcting for finite momentum resolution, muon contamination, and FSI corrections, such distributions should be identical to $N_1^4$ in the absence of additional correlations.
A small difference in $c_4^{\rm QS}$ is observed without the subtraction of such terms~\cite{pubnote:2015}.

\subsubsection{Fits to three-pion correlation functions}
The $2^{\rm nd}$ build technique relies on the extraction of the pair-exchange magnitudes from fits to three-pion correlations.  
We separately fit both the cumulant ($c_3^{\rm QS}$) and full ($C_3^{\rm QS}$) correlations with an Edgeworth parametrization in 3D ($q_{12},q_{13},q_{23}$).
The three-pion correlations and fits are projected onto $Q_3$ for pp, p--Pb, and Pb--Pb collisions in Figs.~\ref{fig:C3fits_pp_K0}-\ref{fig:C3fits_PbPb_K0}. 
The Edgeworth fits have six free parameters, $s,R,\kappa_3$, $\kappa_4$, $\kappa_5$, and $\kappa_6$, \cite{Gangadharan:2015ina} as well as a fixed value of $G$.
In Figs.~\ref{fig:C3fits_pp_K0}-\ref{fig:C3fits_PbPb_K0}, $G=0$.  

\begin{figure}[t!h]
  \centering
  \subfigure[pp]{
    \includegraphics[width=0.48\textwidth]{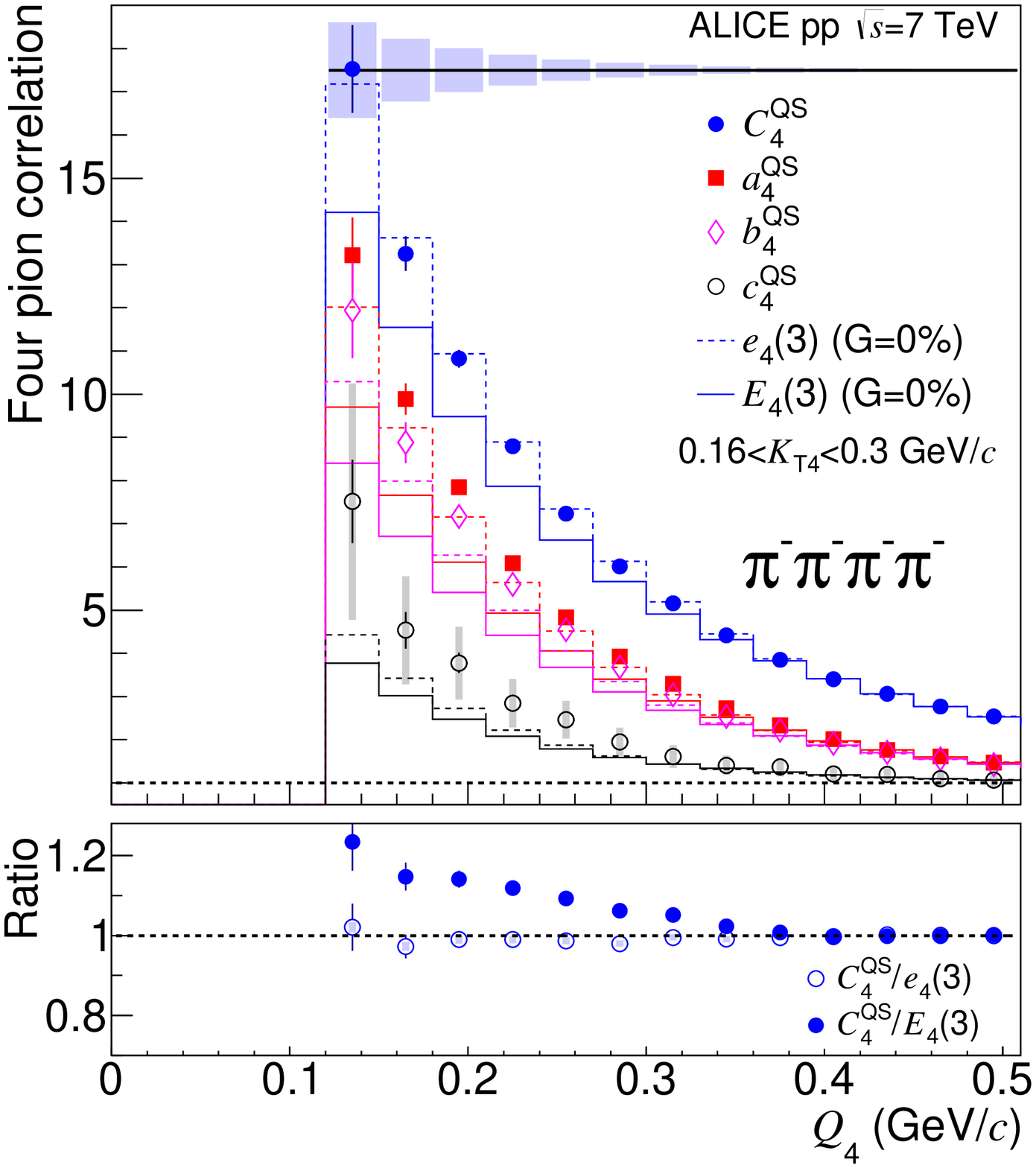}
    \label{fig:C4SC_pp}
  }
  \subfigure[p--Pb]{
    \includegraphics[width=0.48\textwidth]{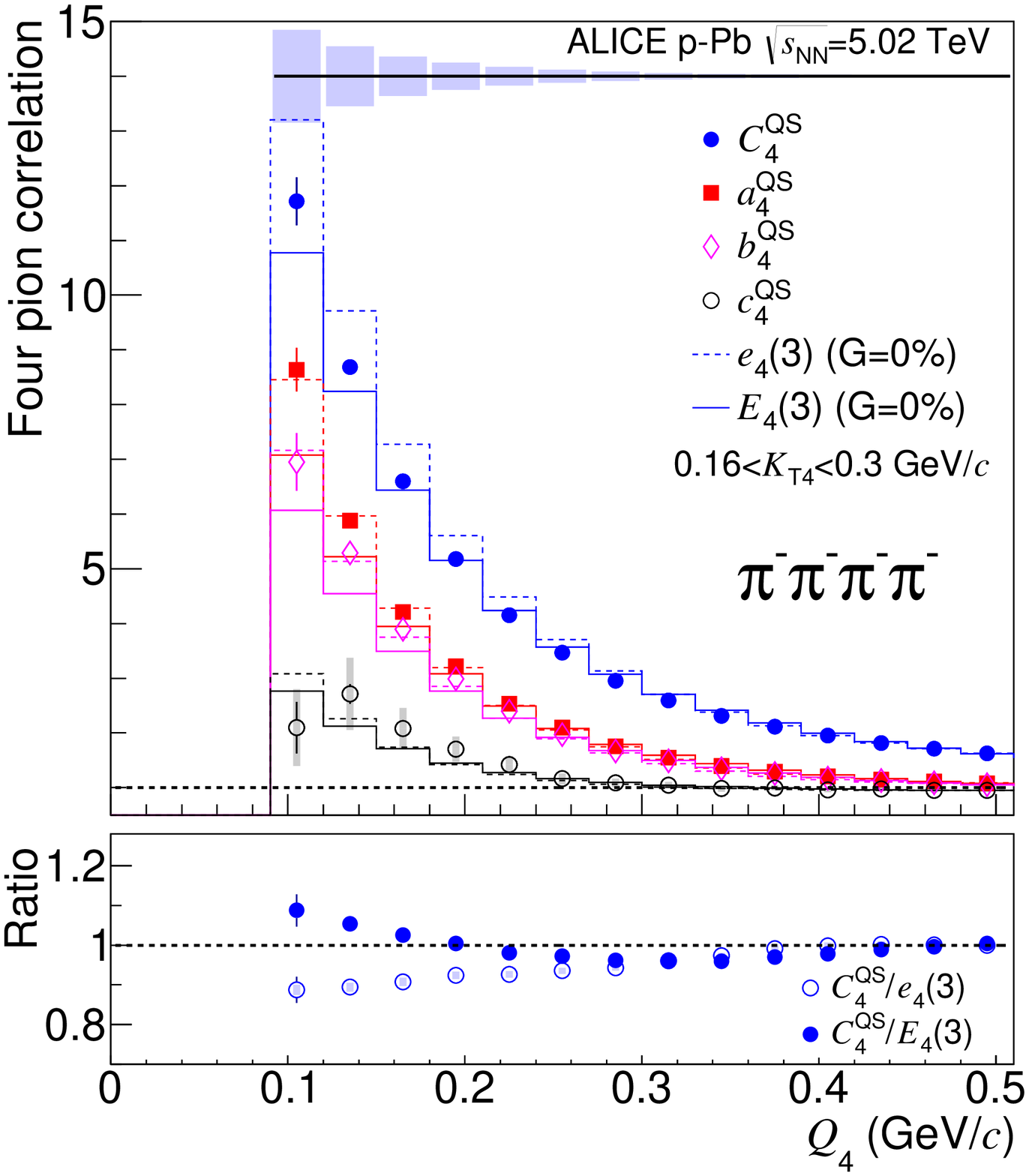}
    \label{fig:C4SC_pPb}
  }
  \subfigure[Pb--Pb]{
    \includegraphics[width=0.48\textwidth]{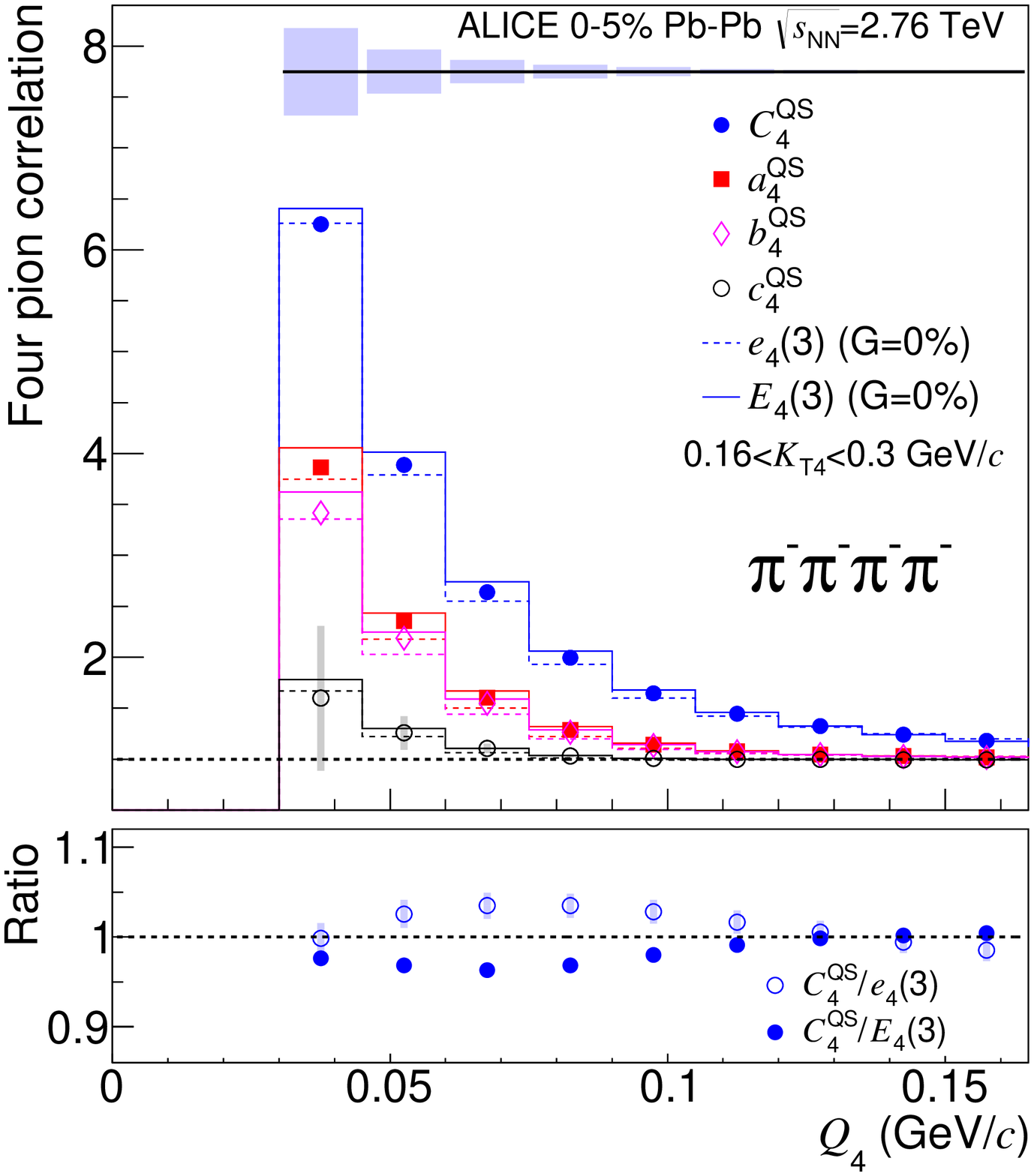}
    \label{fig:C4SC_PbPb_2ndBuild}
  }
  \caption{Same-charge four-pion full ($C_4^{\rm QS}$), partial cumulant ($a_4^{\rm QS}$, $b_4^{\rm QS}$), and cumulant ($c_4^{\rm QS}$) correlations versus $Q_4$ in pp (a), p--Pb (b), and Pb--Pb (c) collisions.  The solid and dashed block histograms represent $E_4(3)$ and $e_4(3)$ with $G=0$, respectively.  Systematic uncertainties shown at the top apply to $C_4^{\rm QS}$.  The systematics for the other correlation functions are obtained by scaling down the shaded band by the relative correlation strengths.  The systematic uncertainties are similar for the expected and measured correlation functions for which the small difference is shown in the ratio. An additional systematic is drawn for $c_4^{\rm QS}$ and is explained in the systematics section.  The bottom panel shows the ratio of measured to expected $C_4^{\rm QS}$. The average of the charge conjugated correlation functions is shown.}
\end{figure}

\subsection{Same-charge three- and four-pion QS correlations}
Figures \ref{fig:C4SC_pp}-\ref{fig:C4SC_PbPb_2ndBuild} present same-charge four-pion correlations in all three collision systems.
Each symmetrization sequence is clearly visible.
Two different expectations are shown: $E_4(3)$ and $e_4(3)$. 
The expected correlations in pp and p--Pb are typically within $10\%$ of measured correlations while being closer, $5\%$, in Pb--Pb.  

\begin{figure}[!h]
  \centering
  \subfigure[$0.16<\KTThree<0.3$ GeV/$c$]{
    \includegraphics[width=0.48\textwidth]{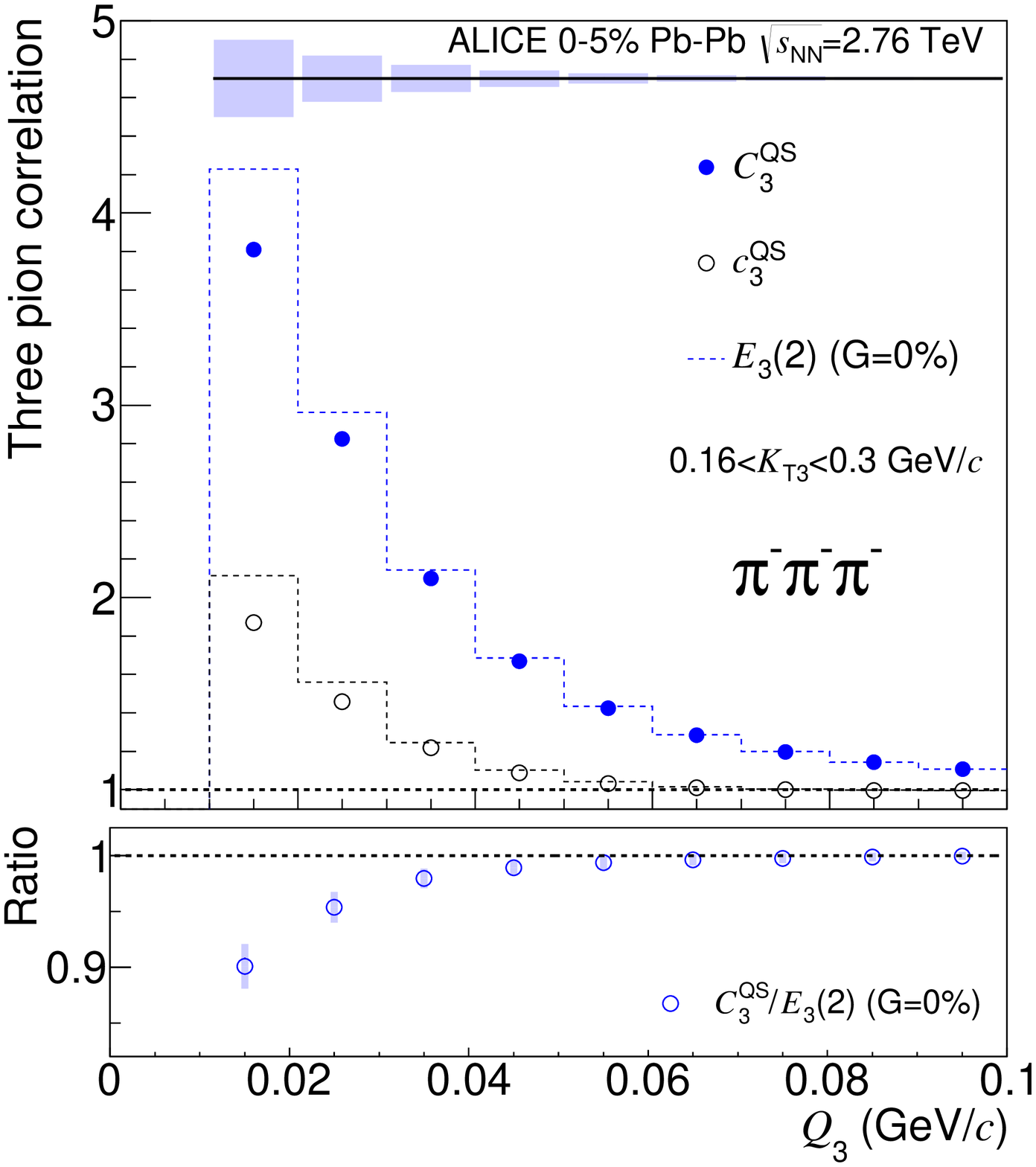}
    \label{fig:C3SC_PbPb_long_K0}
  }
  \subfigure[$0.3<\KTThree<1.0$ GeV/$c$]{
    \includegraphics[width=0.48\textwidth]{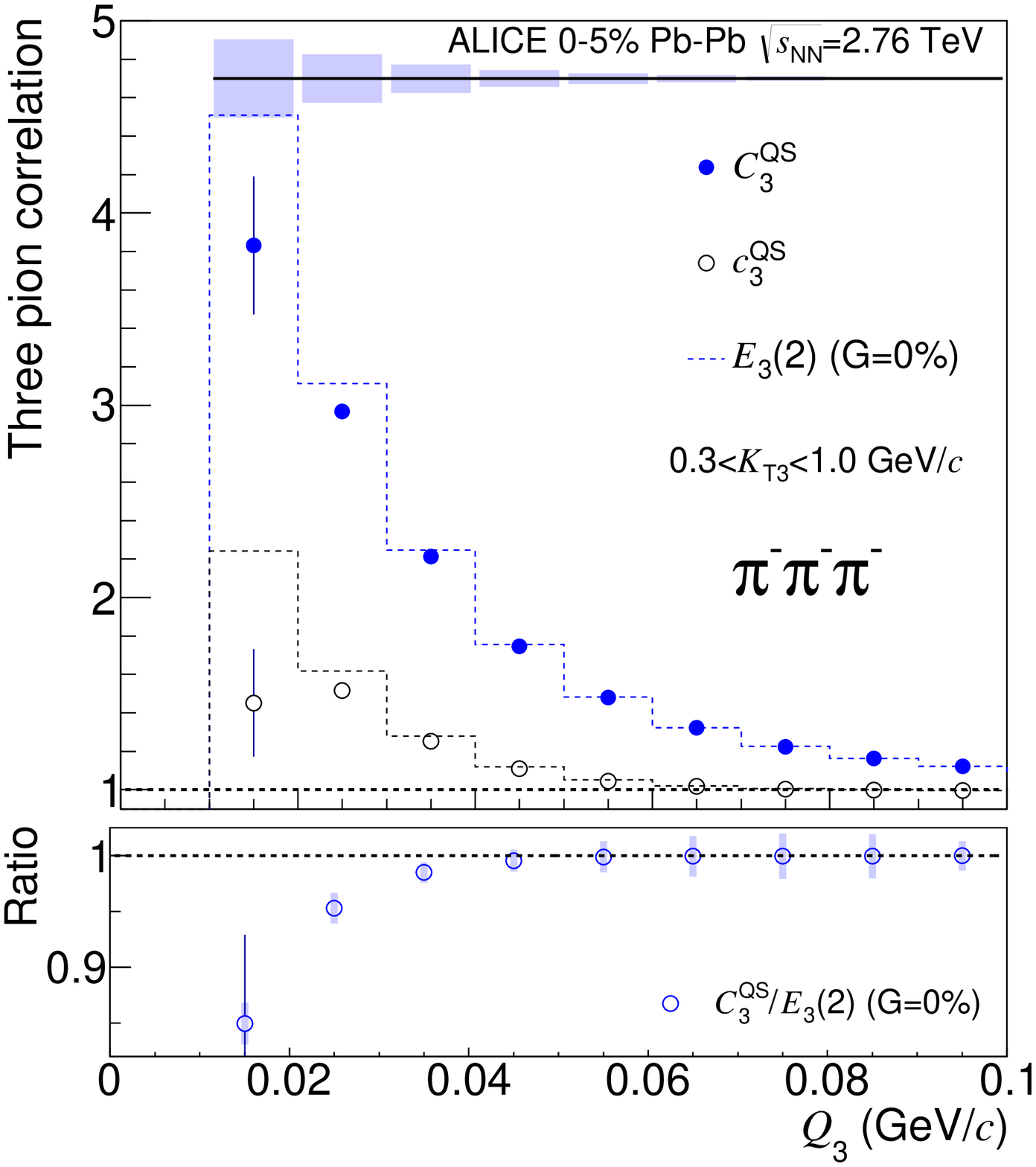}
    \label{fig:C3SC_PbPb_long_K1}
  }
  \caption{Three-pion same-charge full ($C_3^{\rm QS}$) and cumulant ($c_3^{\rm QS}$) correlations versus $Q_3$ in Pb--Pb.  Expected correlations of the $1^{\rm st}$ type are shown with dashed block histograms with $G=0$.  The ratio of measured to expected $C_3^{\rm QS}$ is shown in the bottom panel.  The systematic uncertainties are shown by the shaded bands at the top of the figure as explained in Fig.~\ref{fig:C4SC_pp}.  The average of the charge conjugated correlation functions is shown.}
\end{figure}

Three-pion measured and expected correlations in Pb--Pb are presented in Figs.~\ref{fig:C3SC_PbPb_long_K0}-\ref{fig:C3SC_PbPb_long_K1} for low and high \KTThree.
The expected correlations are of the $1^{\rm st}$ type and assume $G=0$.
The top panels show the full and cumulant three-pion correlations while the bottom panels present the ratio of measured to expected full three-pion correlations.
From the bottom panels we observe a $Q_3$ dependent suppression of measured correlations, compared to the expected correlations.

\begin{figure}[!h]
  \centering
  \subfigure[$0.16<\KTFour<0.3$ GeV/$c$]{
    \includegraphics[width=0.48\textwidth]{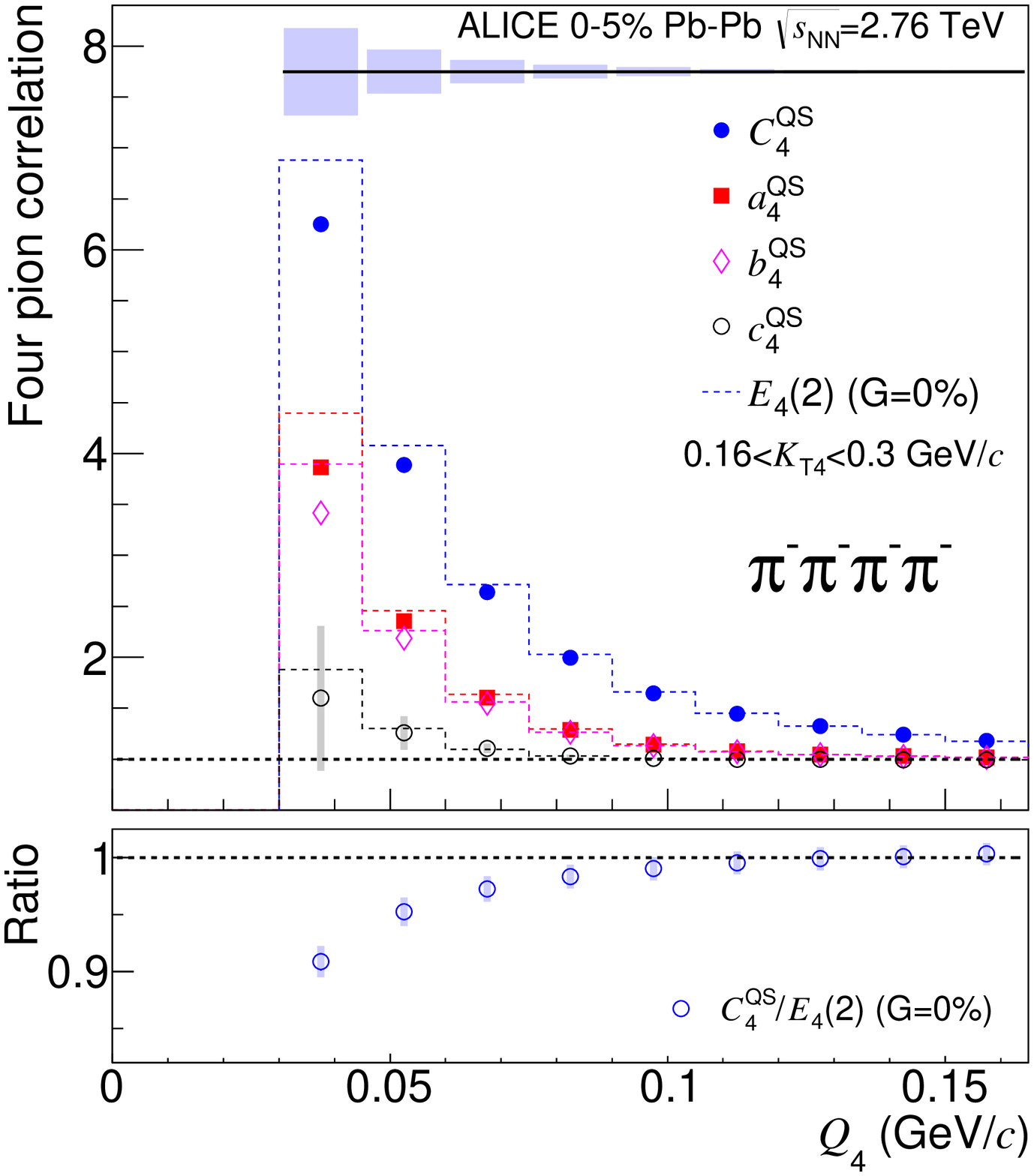}
    \label{fig:C4SC_PbPb_long_K0}
  }
  \subfigure[$0.3<\KTFour<1.0$ GeV/$c$]{
    \includegraphics[width=0.48\textwidth]{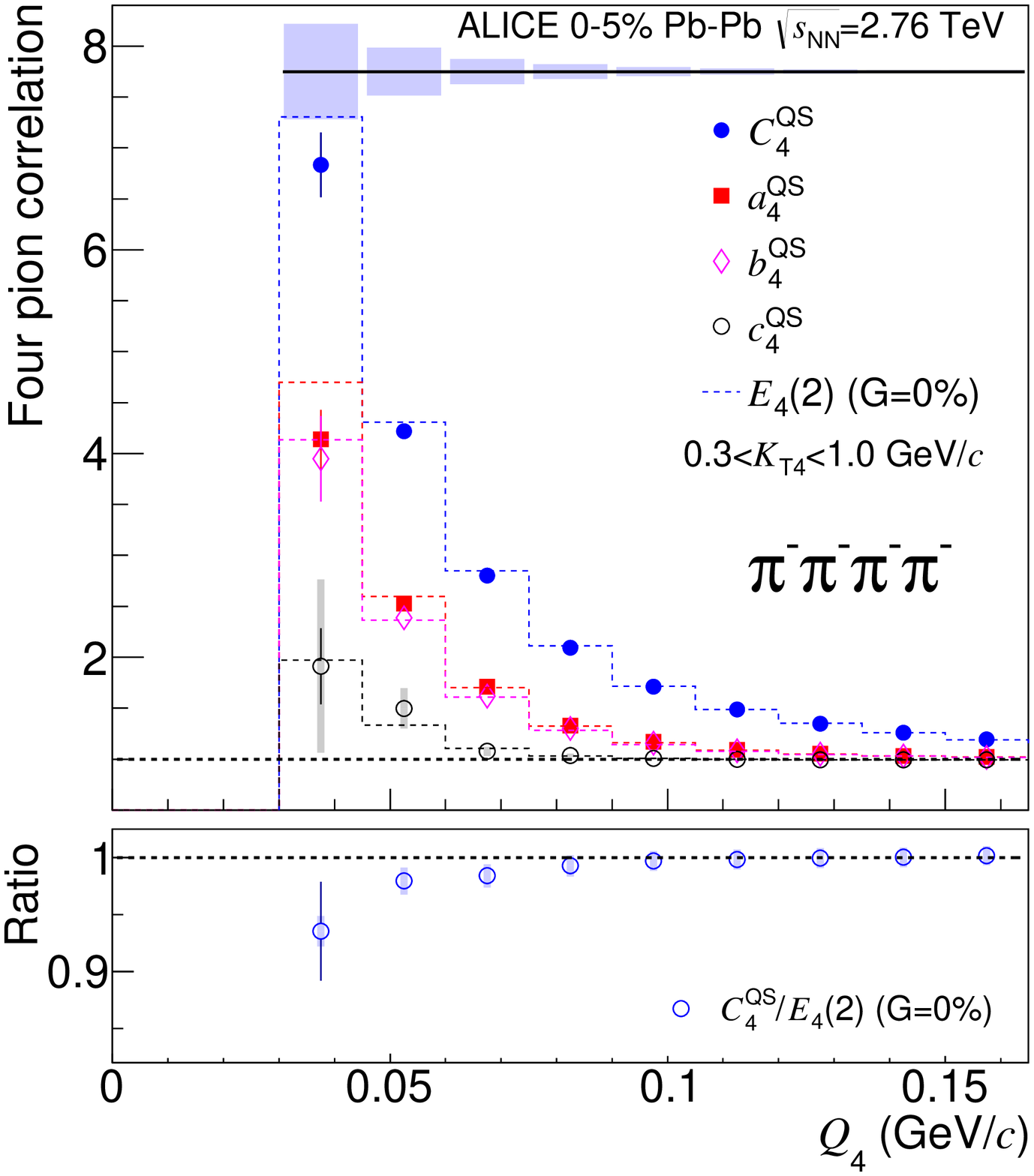}
    \label{fig:C4SC_PbPb_long_K1}
  }
  \caption{Four-pion same-charge full and cumulant correlations versus $Q_4$ in Pb--Pb.  Expected correlations of the $1^{\rm st}$ type are shown with dashed block histograms with $G=0$.  The other details are the same as Fig.~\ref{fig:C4SC_pp}.}
\end{figure}

Four-pion measured correlations are compared to the $E_4(2)$ expectations in Pb--Pb in Figs.~\ref{fig:C4SC_PbPb_long_K0}-\ref{fig:C4SC_PbPb_long_K1} for low and high \KTFour.
Similar to the three-pion case, we observe a $Q_4$ dependent suppression of measured compared to the expected correlations.

\subsection{Extracting a possible coherent fraction}
We now investigate the expected correlations with non-zero values of the coherent fraction, $G$, and compare them to the measured correlations in Pb--Pb.
We use the expected correlations of the $1^{\rm st}$ type to extract the coherent fraction from four-pion correlations.
Owing mostly to limitations of the three-pion fitting procedure, we do not extract the coherent fraction with the $2^{\rm nd}$ type.  
The isospin effect relevant for charged-particle coherent states is neglected in this analysis \cite{Botke:1974ra,Gyulassy:1979yi,Biyajima:1998yh,Akkelin:2001nd}.

\begin{figure}[!h]
  \centering
  \includegraphics[width=0.8\textwidth]{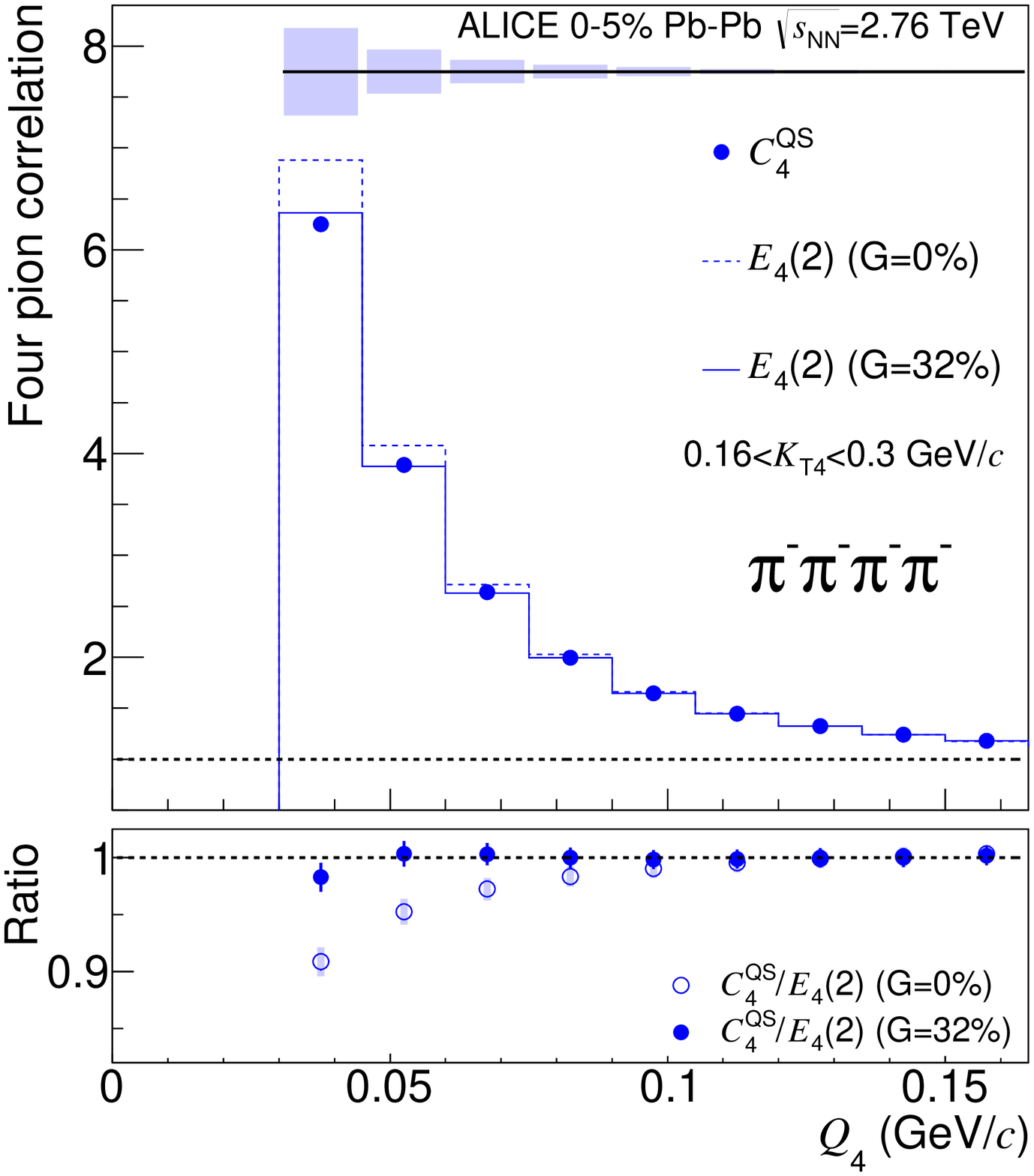}
  \caption{Same-charge four-pion full ($C_4^{\rm QS}$) correlations versus $Q_4$.  Measured and expected correlations of the $1^{\rm st}$ type are shown.  Dashed and solid block histograms show the $G=0$ and $G=32\%$ expected correlations, respectively.  Systematic uncertainties are shown at the top.  The bottom panel shows the ratio of measured to the expected $C_4^{\rm QS}$.  The systematic uncertainties on the ratio are shown with a shaded blue band ($G=0$) and with a thick blue line ($G=32\%$).}
  \label{fig:C4SC_short_K0}
\end{figure}

\begin{figure}[!h]
  \centering
  \includegraphics[width=0.8\textwidth]{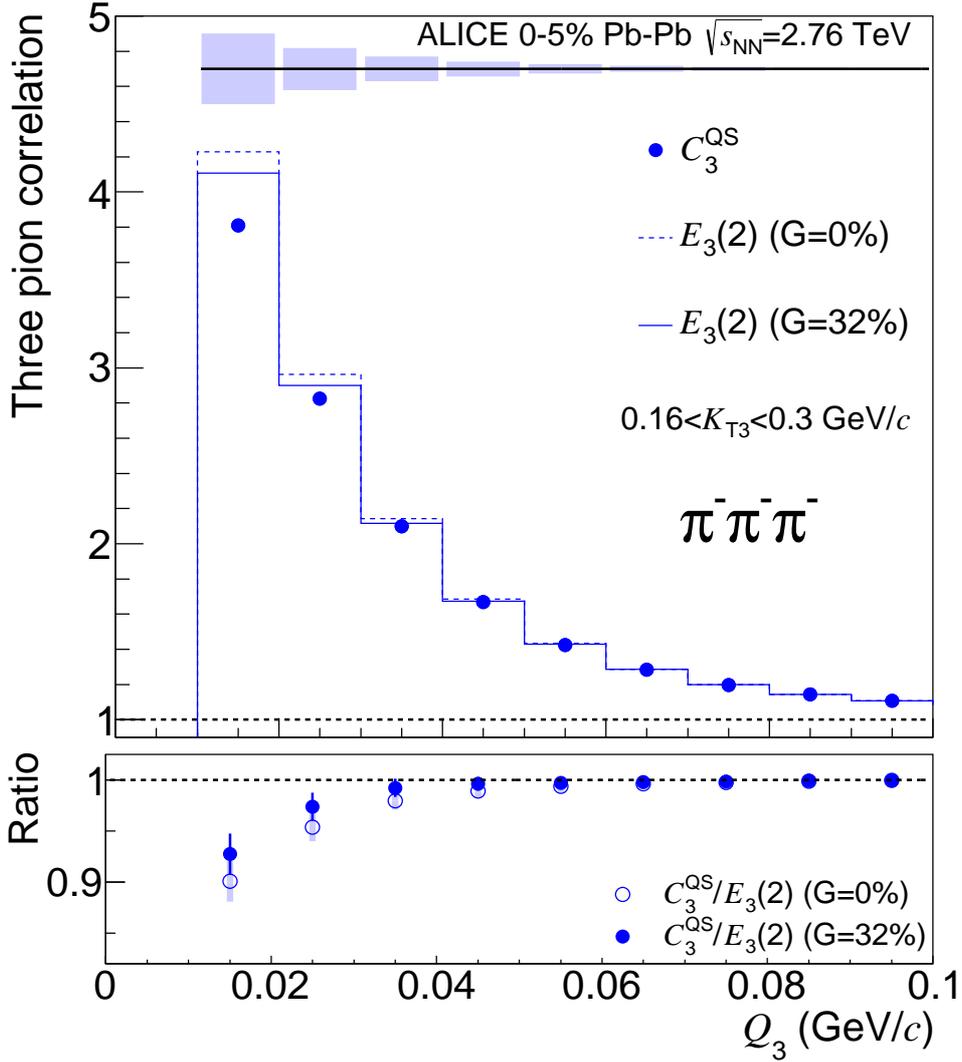}
  \caption{Same-charge three-pion full ($C_3^{\rm QS}$) correlations versus $Q_3$.  Measured and expected correlations of the $1^{\rm st}$ type are shown.  Dashed and solid block histograms show the $G=0$ and $G=32\%$ expected correlations, respectively.  The other details are the same as in Fig.~\ref{fig:C4SC_short_K0}.}
  \label{fig:C3SC_short_K0}
\end{figure}

Figure \ref{fig:C4SC_short_K0} presents same-charge four-pion correlations in Pb--Pb versus $Q_4$ at low \KTFour. 
We observe that the suppression can be partially explained assuming $G=32\%$ which minimizes the $\chi^2$ of the difference of the ratio from unity for $Q_4<0.105$ GeV/$c$.
The $\chi^2$/DOF of the minimum is quite low, 0.34, and is due to the inclusion of high $Q_4$ data in the calculation and the rapidly decreasing QS correlation with $Q_4$.
In Fig.~\ref{fig:C3SC_short_K0} we present same-charge three-pion correlations in Pb--Pb versus $Q_3$ at low \KTThree. 
In contrast to the four-pion case, the value of $G=32\%$ does not satisfactorily explain the suppression.

\begin{figure}[!h]
  \centering
  \subfigure[$0.16<\KTFour<0.3$ GeV/$c$]{
    \includegraphics[width=0.48\textwidth]{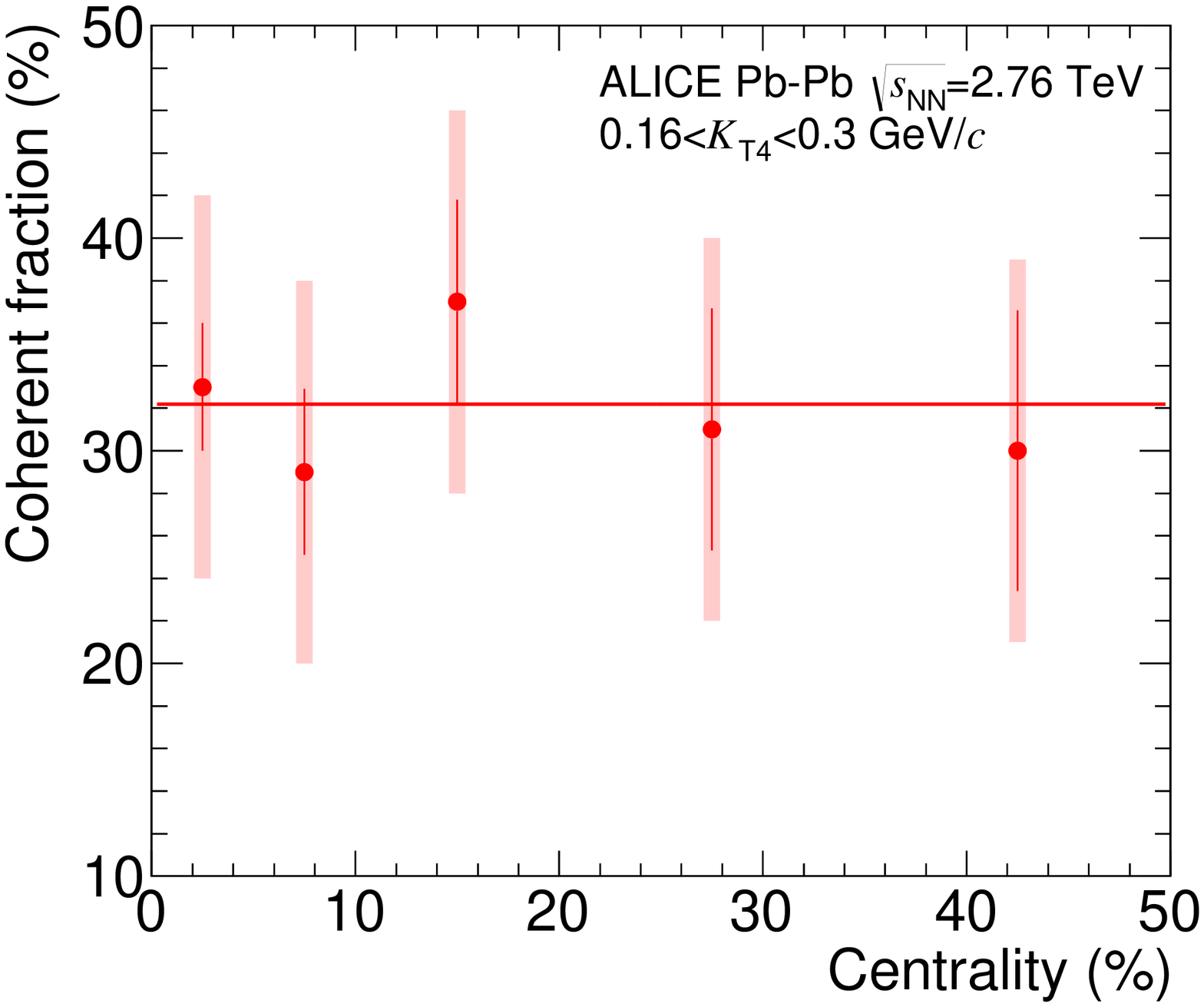}
    \label{fig:GversusCent_K0}
  }
  \subfigure[$0.3<\KTFour<1.0$ GeV/$c$]{
    \includegraphics[width=0.48\textwidth]{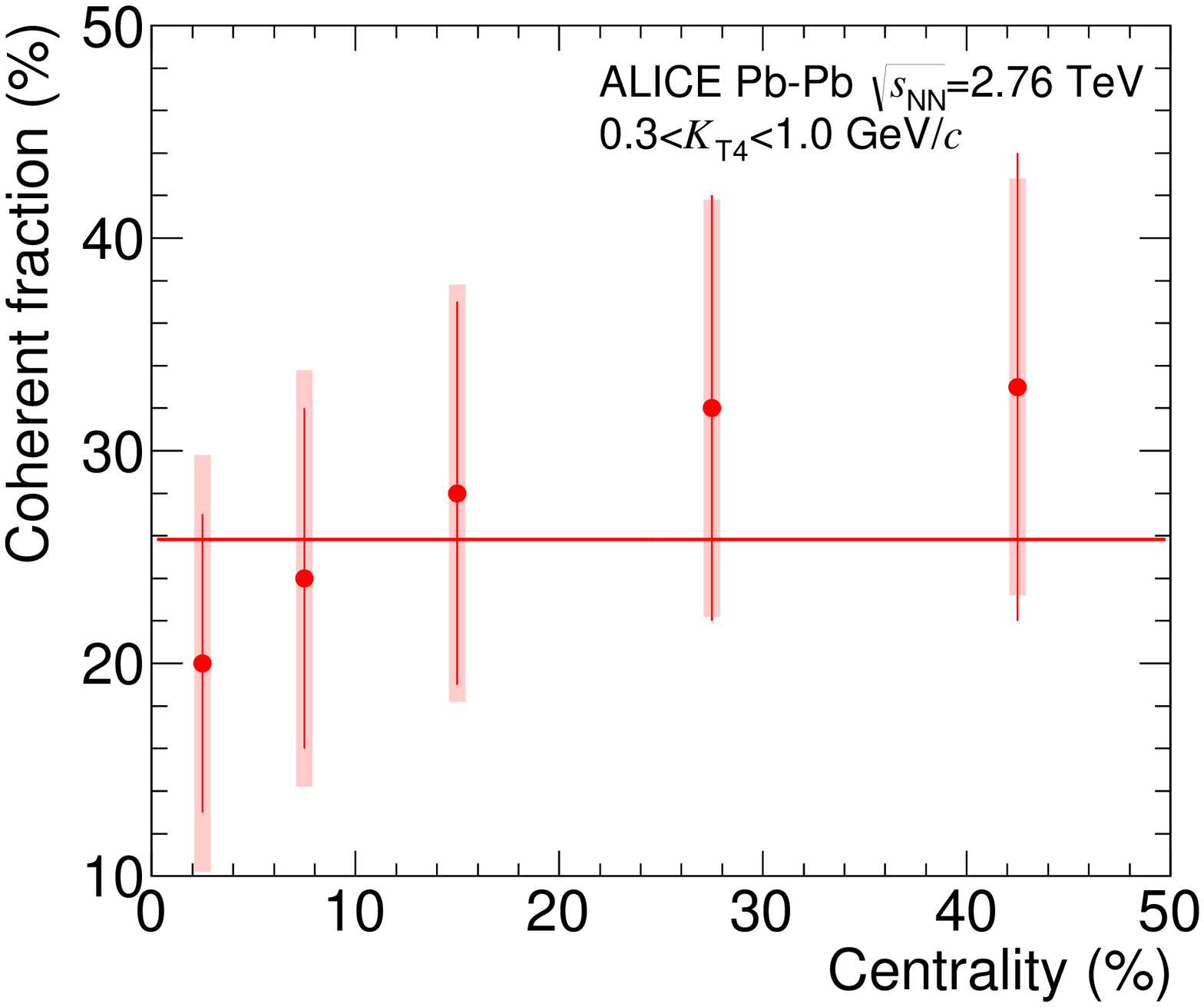}
    \label{fig:GversusCent_K1}
  }
  \caption{The extracted coherent fractions ($G$) from same-charge four-pion correlations versus centrality.  Systematic uncertainties are given by the shaded band.  A linear fit using only the statistical uncertainties is shown by the horizontal red line.}
\end{figure}

We also studied the centrality dependence of the suppression in Pb--Pb.
Figures \ref{fig:GversusCent_K0} and \ref{fig:GversusCent_K1} show the centrality dependence of the extracted coherent fraction for low and high \KTFour.
Within statistical and systematic uncertainties, the coherent fractions are consistent for each centrality interval. 
We also parametrized the coherent component as a point source as opposed to the equal radii assumption used by default.
The point source approximation may be expected to be more appropriate for gluon or pion condensate formation.
The extracted coherent fractions with the point source approximation are shown in a separate note \cite{pubnote:2015}.

Previously \cite{Abelev:2013pqa}, the coherent fractions were extracted from the $r_3$ observable which is intended to isolate the phase of three-pion correlations \cite{Heinz:1997mr,Heinz:2004pv}.  
In contrast to the previous analysis, we estimate $G$ by averaging the suppression in several $Q_3$ or $Q_4$ bins instead of extrapolating $r_3$ to the unmeasured intercept.
This approach was chosen due to the largely flat relative momentum dependence of previous $r_3$ measurements \cite{Abelev:2013pqa,Gangadharan:2015ina}.
The values of $G$ are obtained by averaging the bin-by-bin values within $0.03<Q_4<0.105$ GeV/$c$.
Furthermore, our past analysis did not employ interpolation corrections which are relevant for the expected correlations.  
Correcting for the interpolation biases is expected to lower $r_3$ \cite{Gangadharan:2015ina}.

We extracted coherent fractions in Pb--Pb using the expected correlations of the $1^{\rm st}$ type. 
The expected correlations of the $2^{\rm nd}$ type were shown in all three collision systems but are expected to be less accurate due to more limited dimensionality and the fitting procedure of three-pion correlations.
Being such, we could not reliably extract a value of $G$ with the $2^{\rm nd}$ build technique.
The $2^{\rm nd}$ type is, however, preferred in low multiplicity events, where non-negligible background correlations exist.

One of the most commonly cited sources of coherent pion emission is the DCC \cite{Bjorken:1993cz,Bjorken:1997re}, which may occur as a consequence of chiral symmetry restoration.
The most common prediction of the DCC is the fluctuation of charged to neutral pion production at low $\pt$.
If a single DCC domain is created within each event, we may expect a surplus of coherent charged pions in one event, while in another event, only coherent neutral pions are present.  
We investigated this possibility by first isolating a narrow multiplicity class at higher $\pt$, $0.35<\pt<0.5$ GeV/$c$, within the 0-5$\%$ centrality class determined with the V0 detectors.
From the multiplicity distribution of charged pions at the higher \pt~interval, we retain events which were within 1 standard deviation from the mean of the distribution.  
We then analyzed the multiplicity distribution of charged pions at low \pt, $0.16<\pt<0.25$ GeV/$c$.  
Events with low \pt~multiplicities below the mean of the distribution were stored separately from those events above the mean.  
We do not observe a significant change of the suppression for events below or above the mean.  
The finding disfavors single-domain DCCs but does not rule out multidomain DCCs, for which independently coherent charged and neutral pions may be found in a single event~\cite{Bjorken:1993cz,Bjorken:1997re}.

%
%
%
\section{Systematic uncertainties}
\label{sec:systematics}
We consider several sources of systematic uncertainty pertaining to the methodology and finite detector resolution.
Below we describe each systematic uncertainty studied in order of decreasing magnitude.
Some systematic uncertainties apply to only measured or expected correlations while others apply to both.
The given values of the uncertainties apply to four-pion correlations.  
The values for three-pion correlations are generally smaller.
\begin{enumerate}
\item {\it $f_c$ scale}.  The fraction of pion tracks from short-lived emitters for which QS and FSI correlations are experimentally observable is quantified with the $f_c$ parameter.  
From previous studies in ALICE using fits to $\pi^+\pi^-$ FSI correlations, we estimate that $f_c=0.84\pm0.03$ \cite{Abelev:2013pqa}.  
We vary $f_c$ within its uncertainties from the previous analysis.  
The uncertainty derived from varying $f_c$ applies to both measured and expected correlations and is about $6\%$ at low $Q_4$.
As the uncertainty on $f_c$ given here does not account for the assumption of a universal $f_c$ for both QS and Coulomb correlations (see discussion in Section 3), we have also considered more extreme variations given by $f_c=0.63$ and $f_c=0.92$.
The systematic variations of measured and expected correlations are largely correlated.
With $f_c=0.63$, the ratio of measured to expected four-pion correlations increased by about $2\%$ at low $Q_4$ as compared to the ratio formed with our default $f_c=0.84$.
\item {\it FSI variation}.  The default two-pion FSI correlation $K_2$, together with the default value $f_c=0.84$, gives a satisfactory description of $\pi^+\pi^-$ correlations \cite{Abelev:2013pqa}. 
We find that increasing the FSI correlation strength, $|K_2-1|$, by $5\%$ while decreasing $f_c$ to 0.806 also provides a satisfactory description of $\pi^+\pi^-$ correlations.  
The analysis was redone with such modifications, and the ratio of measured to expected four-pion correlations changed by less than $0.5\%$.
\item {\it $T_{ij}$ extraction at high $q$}.  The $1^{\rm st}$ type of expected correlations use the pair-exchange magnitudes ($T_{ij}$) extracted from two-pion correlations.
The extraction of $T_{ij}$ becomes problematic at large $q$, where the measured two-pion QS correlations fluctuate beneath the baseline due to finite statistics.
For such bins we set $T_{ij}=0$.
We also constructed a separate expected correlation where the entire triplet or quadruplet was skipped if any pair $T_{ij}$ was negative.
Half of the difference between these two builds was assigned as an uncertainty which is about $4\%$ at high $Q_4$ and less than $0.1\%$ at low $Q_4$.
\item {\it Interpolation}.  We apply a data-driven approach to correct for interpolation biases, as already mentioned.  
From studies with different interpolation schemes, we find a $1\%$ systematic uncertainty on the expected correlations at low $Q_4$. 
\item {\it Mid-lived emitters}.  The extraction of the multipion QS correlations from the measured distributions in Eqs.~\ref{eq:N1N1N1N1}-\ref{eq:N4} relies on the $f_{41},f_{42},f_{43},f_{44}$ coefficients in Ref.~\cite{Gangadharan:2015ina}. 
The default values were derived in the ``core-halo" picture of particle production, for which there are only short and long-lived emitters.  
In general there are also mid-lived emitters (e.g.\ $\omega$ decays) which modify the $f$ coefficients and can be estimated using the \textsc{therminator} model.  
The effect was found to be quite small \cite{Gangadharan:2015ina} and leads to a $0.5\%$ uncertainty at high $Q_4$.
\item {\it Renormalization}.  To account for small normalization differences between two-, three-, and four-pion correlation functions, the expected correlations are re-normalized to the ones measured at high $Q_4$.
In central Pb--Pb, the renormalizations are about 0.9997 ($E_3(2)$), 1.005 ($E_4(2)$), and 1.07 ($e_4(3)$).
The interval in Pb--Pb is $0.125<Q_4<0.145$ GeV/$c$ in central collisions and varies smoothly to $0.165<Q_4<0.185$ GeV/$c$ in peripheral collisions.
The interval in pp and p--Pb is $0.46<Q_4<0.49$ GeV/$c$.
We take an interval shifted by 15 (60) MeV/$c$ in Pb--Pb (pp and p--Pb).
\item {\it Detector resolution}.  Numerous effects related to finite detector resolution were checked.  
The charge conjugated correlation functions were consistent within statistical uncertainties.  
Similarly, the polarity of the solenoidal magnetic field had a negligible effect on the correlation functions.  
We compared Pb--Pb data from two different data-taking periods which were known to have different tracking efficiencies.  
The measured and expected correlation functions differed by less than $0.5\%$.
Finite momentum resolution is known to smear the correlation functions, decreasing the correlation strength at low relative momentum for all orders of correlation functions.  
We correct for finite momentum resolution using HIJING (Pb--Pb) and PYTHIA\cite{Sjostrand:2006za} (pp and p--Pb) data simulated with the ALICE detector response.  
The uncertainty on the momentum resolution at low \pt~is governed by the material budget uncertainty of the ALICE detector and is estimated to be less than $10\%$.  
The corresponding uncertainty on the measured and expected correlations is about $1\%$.
Our pion purity is estimated to be about $96\%$ for which the remaining $4\%$ impurity is dominated by muon contamination.  
Simulations have shown that most of the muons in our sample originate from charged pion decays for which QS and FSI correlations are expected with primary pions.  
We apply muon corrections similar to Refs.~\cite{Abelev:2013pqa,Abelev:2014pja}.
We assign a $2\%$ uncertainty to the muon correction procedure.
The tracking efficiency of the ALICE detector decreases rapidly for $\pt<0.2$ GeV/$c$~\cite{Abelev:2014ffa}.  
To estimate the potential bias caused by the tracking efficiency, we randomly discard pions in \textsc{therminator} according to the TPC reconstruction efficiency.
We do not observe a bias on the measured nor expected correlation functions which could cause an artificial suppression.
\end{enumerate}
In addition to the above mentioned sources of systematics, we also applied an additional uncertainty to cumulant correlation functions, $c_4^{\rm QS}$.
The cumulant correlations were found to be much more sensitive to effects induced by low statistics at low $Q_4$.
The additional uncertainty is several tens of percents for the lowest $Q_4$ bin.

Most of the systematic uncertainties were found to be similar in magnitude and highly correlated for both measured and expected correlations.  
As a consequence, the systematics largely cancel in the ratio of measured to expected.
For the ratio, we apply the maximum difference of measured and expected systematics.
The systematic uncertainties for the ratio are dominated by the interpolator and mid-lived emitter uncertainty at low $Q_4$.
At high $Q_4$, the muon corrections and the extraction of $T_{ij}$ at high $q$ dominate the uncertainties.

%
%
%
\section{Possible origins of the suppression}
\label{sec:origins}
A suppression of three- and four-pion Bose-Einstein correlations compared to the expectations from two-pion measurements has been observed in Pb--Pb collisions.
Below we list our investigations into the origin of the suppression.

\begin{enumerate}
\item {\it Quantum coherence}.  Incorporating the effects of quantum coherence can perhaps explain the four-pion suppression in Fig.~\ref{fig:C4SC_short_K0} with a centrality averaged coherent fraction of $32\% \pm 3\%$(stat) $\pm 9\%$(syst).  However, the same coherent fraction fails to explain the suppression at the three-pion level in Fig.~\ref{fig:C3SC_short_K0}.  In particular, the suppression at the lowest $Q_3$ and $Q_4$ intervals cannot be resolved with the same coherent fraction as needed at higher $Q_3$ and $Q_4$ intervals.
The isospin effect for charged-pion coherent states \cite{Botke:1974ra,Gyulassy:1979yi,Biyajima:1998yh,Akkelin:2001nd} has not been calculated, since the expressions which incorporate isospin conservation do not exist at the four-pion level.
For $G=32\%$, the isospin effect increases the intercept of two- and three-pion correlations by about $1\%$ and $3\%$, respectively.
The effect on the expected correlations at finite relative momentum has not been calculated.
\item {\it Coulomb repulsion}.  Same-charge pions experience Coulomb and strong repulsion which is stronger for quadruplets than for pairs.  
The four-pion Coulomb corrections used in this analysis correspond to the asymptotic limit of the Coulomb wave function as mentioned before.  
Previous studies \cite{Alt:1998nr} have justified the use of such wave functions for the characteristic freeze-out volumes and relative momenta studied in this analysis.
We have also shown that the cumulant ($c_4^{\rm QS}$) of mixed-charge correlations are near unity after FSI corrections. 
In the case that the genuine multipion Coulomb interactions are not negligible, we modify the three- and four-pion FSI correlations by an amount, $x$, needed to resolve the suppression (residue) of same-charge (mixed-charge) correlations.
The FSI factors are modified as: $K_{3,4} \rightarrow x|K_{3,4}-1|+1$.
\begin{table}
  \center
  \begin{tabular}{| c | c | c | c | c | c |}
    \hline
    & $--+$ & $+++$ & $---+$ & $--++$ & $++++$ \\ \hline
    Low \KTThree,\KTFour & $1.07\pm0.01$ & $1.16\pm0.02$ & $1.6\pm0.1$ & $0.89\pm0.02$ & $1.17\pm0.02$  \\ \hline
    High \KTThree,\KTFour & $1.06\pm0.01$ & $1.13\pm0.02$ & $1.2\pm0.1$ & $0.89\pm0.02$ &  $1.09\pm0.02$ \\ \hline
  \end{tabular}
  \caption{The $x$ factors used to modify the multipion FSI factor such that the suppression of same-charge correlations and the residues of mixed-charged cumulants are resolved.  The multipion FSI factor is modified according to: $K_{3,4} \rightarrow x|K_{3,4}-1|+1$.  With $---+$ correlations, only $K_4$ was modified and not $K_3$ which is also used to isolate the cumulant.  We further note that $x$ is more $Q_3$ and $Q_4$ dependent for the case of $+++$ and $++++$.  We find that for the lowest $Q_3$ and $Q_4$ bin, $x$ is about 1.2.}
  \label{tab:xFactors}
\end{table}
The $x$ factors given in Tab.~\ref{tab:xFactors} demonstrate that if the suppression is solely caused by genuine multipion Coulomb effects, they should modify the two-body approximation by up to $20\%$ at low relative momentum for the case of same-charge three- and four-pion correlations.
Such large multibody Coulomb correlations are not expected from the arguments provided in Ref.~\cite{Alt:1998nr}.
\item{\it Mid-lived emitters}.  Uncertainties of mid-lived resonance production ($\Gamma\sim 10$ MeV) result in uncertainties of $f_{44}$, $f_{43}$, $f_{42}$, and $f_{41}$ \cite{Gangadharan:2015ina} which are used to isolate the QS correlations from the measured distributions.  We investigated the possibility of decreasing $f_{44}$ while equally increasing $f_{41}$, $6f_{42}$, and $4f_{43}$ following the unitary probability constraint: $f_{44}+4f_{43}+6f_{42}+f_{41}=1$.
Decreasing $f_{44}$ by 0.08 resolves the suppression for $Q_4<0.06$ while 0.04 is more appropriate for larger $Q_4$. 
However, as a consequence the $---+$ and $--++$ cumulant correlations increase by as much as 0.2 at low $Q_4$, which leaves larger unexplained residues.
\item {\it Background correlations}.  Event generators such as HIJING and AMPT \cite{Lin:2004en} do not include the effects of QS nor FSI and may thus be used to estimate background correlations.
We checked two-, three-, and four-pion correlation functions in the $5\%$ most central events from HIJING and AMPT.  
All orders of correlation functions were consistent with unity.
\item {\it Multipion phases}.  The expected correlations ignore the three- and four-pion Fourier transform phases \cite{Heinz:1997mr}.  
The $r_3$ observable was extracted in ALICE \cite{Abelev:2013pqa} and \textsc{therminator} \cite{Gangadharan:2015ina} and no significant $Q_3$ dependence was found.  
As the trend of $G$ with $Q_3$ and $Q_4$ is opposite to that expected from the phases \cite{pubnote:2015}, we find them unlikely to explain the suppression.
\item {\it Multipion distortions}.  At high freeze-out phase-space density, all higher order symmetrizations, which are usually neglected, can contribute significantly to all orders of correlation functions \cite{Zajc:1986sq,Pratt:1993uy,Pratt:1994cg,Lednicky:1999xz,Zhang:1998sz}.
The distortions have been calculated for two-pion correlations and recently for three- and four-pion correlations \cite{Gangadharan:2015ina}.
The calculations suggest that the ratio of measured to expected correlations is robust with respect to this effect.
\end{enumerate}

%
%
%
\section{Summary}
\label{sec:summary}
Three- and four-pion QS correlations have been measured in pp, p--Pb, and Pb--Pb collisions at the LHC.
The measured same-charge multipion correlations are compared to the expectation from lower order experimental correlation functions.
A significant suppression of multipion Bose-Einstein correlations has been observed in Pb--Pb collisions.
The ratio of measured to expected same-charge four-pion correlations is about $6\sigma$ below unity in our lowest $Q_4$ interval.

In pp and p--Pb collisions, owing to background correlations at low multiplicity in two-pion correlation functions, we compare the measured four-pion correlations to the expectation from fits to three-pion correlations ($E_4(3)$ and $e_4(3)$).
Three-pion correlation functions contain substantially larger QS correlations and reduced background correlations, which makes them a preferred base for higher order expectations in pp and p--Pb collisions.
We do not observe a significant suppression of four-pion correlations in pp nor p--Pb collisions.
However, the more limited dimensionality and fitting procedure to three-pion correlations makes $E_4(3)$ and $e_4(3)$ expectations less accurate than $E_4(2)$. 
Nevertheless, despite the presence of the non-femtoscopic background, we also performed the analysis in pp and p--Pb collisions with the first type of expected correlations ($E_4(2)$ and $E_3(2)$).
No significant suppression was observed in pp or p--Pb collisions, although the unknown strength of the non-femtoscopic background prevents an absolute statement.

Mixed-charge four-pion correlations have also been measured.  
They are used to demonstrate the effectiveness of the cumulant isolation via the event-mixing techniques as well as that of the FSI, muon, and momentum resolution corrections.  
The mixed-charge cumulant correlations are shown to be near unity although a finite residue exists with both types of mixed-charge correlations.

The suppression of same-charge three- and four-pion correlations in Fig.~\ref{fig:C4SC_short_K0} and \ref{fig:C3SC_short_K0} cannot be unambiguously resolved with any of the possible origins discussed. 
For example, if genuine multipion Coulomb interactions are non negligible, a large increase of as much as $20$\% beyond the two-body approximation would be needed to account for the observed suppression.
On the other hand, a coherent fraction of about $32\% \pm 3\%$(stat) $\pm 9\%$(syst) could largely explain the four-pion suppression, but the same value cannot explain the three-pion suppression.
There does not appear to be a significant centrality dependence to the extracted coherent fractions.
The weak \kT~dependence of the coherent fractions does not favor the formation of Bose-Einstein condensates nor disoriented chiral condensates, which are expected to radiate mostly at low $\pt$.
The suppression observed in this analysis appears to extend at least up to $\pt\sim340$ MeV/$c$.


\ifpreprint
\iffull
\newenvironment{acknowledgement}{\relax}{\relax}
\begin{acknowledgement}
\section*{Acknowledgements}
We would like to thank Richard Lednick{\' y} and Tam{\' a}s Cs{\" o}rg{\H o} for numerous helpful discussions.

The ALICE Collaboration would like to thank all its engineers and technicians for their invaluable contributions to the construction of the experiment and the CERN accelerator teams for the outstanding performance of the LHC complex.
The ALICE Collaboration gratefully acknowledges the resources and support provided by all Grid centres and the Worldwide LHC Computing Grid (WLCG) collaboration.
The ALICE Collaboration acknowledges the following funding agencies for their support in building and
running the ALICE detector:
State Committee of Science,  World Federation of Scientists (WFS)
and Swiss Fonds Kidagan, Armenia;
Conselho Nacional de Desenvolvimento Cient\'{\i}fico e Tecnol\'{o}gico (CNPq), Financiadora de Estudos e Projetos (FINEP),
Funda\c{c}\~{a}o de Amparo \`{a} Pesquisa do Estado de S\~{a}o Paulo (FAPESP);
National Natural Science Foundation of China (NSFC), the Chinese Ministry of Education (CMOE)
and the Ministry of Science and Technology of China (MSTC);
Ministry of Education and Youth of the Czech Republic;
Danish Natural Science Research Council, the Carlsberg Foundation and the Danish National Research Foundation;
The European Research Council under the European Community's Seventh Framework Programme;
Helsinki Institute of Physics and the Academy of Finland;
French CNRS-IN2P3, the `Region Pays de Loire', `Region Alsace', `Region Auvergne' and CEA, France;
German Bundesministerium fur Bildung, Wissenschaft, Forschung und Technologie (BMBF) and the Helmholtz Association;
General Secretariat for Research and Technology, Ministry of Development, Greece;
National Research, Development and Innovation Office (NKFIH), Hungary;
Department of Atomic Energy and Department of Science and Technology of the Government of India;
Istituto Nazionale di Fisica Nucleare (INFN) and Centro Fermi -
Museo Storico della Fisica e Centro Studi e Ricerche ``Enrico Fermi'', Italy;
Japan Society for the Promotion of Science (JSPS) KAKENHI and MEXT, Japan;
Joint Institute for Nuclear Research, Dubna;
National Research Foundation of Korea (NRF);
Consejo Nacional de Cienca y Tecnologia (CONACYT), Direccion General de Asuntos del Personal Academico(DGAPA), M\'{e}xico, Amerique Latine Formation academique - 
European Commission~(ALFA-EC) and the EPLANET Program~(European Particle Physics Latin American Network);
Stichting voor Fundamenteel Onderzoek der Materie (FOM) and the Nederlandse Organisatie voor Wetenschappelijk Onderzoek (NWO), Netherlands;
Research Council of Norway (NFR);
National Science Centre, Poland;
Ministry of National Education/Institute for Atomic Physics and National Council of Scientific Research in Higher Education~(CNCSI-UEFISCDI), Romania;
Ministry of Education and Science of Russian Federation, Russian
Academy of Sciences, Russian Federal Agency of Atomic Energy,
Russian Federal Agency for Science and Innovations and The Russian
Foundation for Basic Research;
Ministry of Education of Slovakia;
Department of Science and Technology, South Africa;
Centro de Investigaciones Energeticas, Medioambientales y Tecnologicas (CIEMAT), E-Infrastructure shared between Europe and Latin America (EELA), 
Ministerio de Econom\'{i}a y Competitividad (MINECO) of Spain, Xunta de Galicia (Conseller\'{\i}a de Educaci\'{o}n),
Centro de Aplicaciones Tecnológicas y Desarrollo Nuclear (CEA\-DEN), Cubaenerg\'{\i}a, Cuba, and IAEA (International Atomic Energy Agency);
Swedish Research Council (VR) and Knut $\&$ Alice Wallenberg
Foundation (KAW);
Ukraine Ministry of Education and Science;
United Kingdom Science and Technology Facilities Council (STFC);
The United States Department of Energy, the United States National
Science Foundation, the State of Texas, and the State of Ohio;
Ministry of Science, Education and Sports of Croatia and  Unity through Knowledge Fund, Croatia;
Council of Scientific and Industrial Research (CSIR), New Delhi, India;
Pontificia Universidad Cat\'{o}lica del Per\'{u}.
\end{acknowledgement}
\ifbibtex
\bibliographystyle{utphys}
\bibliography{pmain}{}
\else
\input{refpreprint.tex}
\fi
\appendix
\section{Appendix}
\label{app:supp}
Given the experimentally measured two-pion correlation functions, one may build the expectation for higher order correlation functions using the equations of quantum statistics. 
The measured two-pion correlation functions are first corrected for experimental distortions: momentum resolution and muon contamination.  
Corrections for long-lived emitters and FSI are then performed to extract the genuine QS correlation according to $C_2=(1-f_c^2) + f_c^2 K_2 C_2^{\rm QS}$ \cite{Sinyukov:1998fc}.
In the case of no coherent emission, the pair-exchange magnitudes ($T_{ij}$) can be extracted according to: $C_2^{\rm QS} = 1 + T_{ij}^2$.
The extracted pair-exchange magnitudes are then used to build the expectation for higher order QS correlations \cite{Andreev:1992pu,Csorgo:1999sj,Gangadharan:2015ina}.
In the absence of coherent emission and multipion phases, the three- and four-pion expected QS correlations are
\begin{eqnarray}
E_3 &=& 1 + [T_{12}^2 + {\rm c.p.}] \nonumber \\
&+& 2 T_{12}T_{23}T_{31}, \label{eq:C3NoPhase} \\
E_4 &=& 1 + [T_{12}^2 + {\rm c.p.}] \nonumber \\ 
&+& [T_{12}^2T_{34}^2 + {\rm c.p.}] \nonumber \\ 
&+& 2 [T_{12}T_{23}T_{31} + {\rm c.p.}] \nonumber \\ 
&+& 2 [T_{12}T_{23}T_{34}T_{41} + {\rm c.p.}], \label{eq:C4NoPhase} 
\end{eqnarray}
where ${\rm c.p.}$ stands for the cyclically permuted terms.
The equations which include partial coherence can be found in Ref.~\cite{Andreev:1992pu,Csorgo:1999sj,Gangadharan:2015ina}.
The $T_{ij}$ factors are tabulated from the first pass over the data and used to build higher order correlations by means of a weight applied to the fully mixed-event distribution in the second and final pass.

Each symmetrization sequence is formed with a product of pair-exchange magnitudes.  
Single-pair, double-pair, triplet, and quadruplet sequences are represented by $T_{ij}T_{ji}$, $T_{ij}^2T_{kl}^2$, $T_{ij}T_{jk}T_{ki}$, $T_{ij}T_{jk}T_{kl}T_{li}$, respectively.
The sum of the appropriate symmetrization sequences yields the expected versions of $C_4^{\rm QS}$, $a_4^{\rm QS}$, $b_4^{\rm QS}$, and $c_4^{\rm QS}$.
\newpage
\section{The ALICE Collaboration}
\label{app:collab}



\begingroup
\small
\begin{flushleft}
J.~Adam\Irefn{org40}\And
D.~Adamov\'{a}\Irefn{org84}\And
M.M.~Aggarwal\Irefn{org88}\And
G.~Aglieri Rinella\Irefn{org36}\And
M.~Agnello\Irefn{org110}\And
N.~Agrawal\Irefn{org48}\And
Z.~Ahammed\Irefn{org132}\And
S.~Ahmad\Irefn{org19}\And
S.U.~Ahn\Irefn{org68}\And
S.~Aiola\Irefn{org136}
% ...
\renewcommand\labelenumi{\textsuperscript{\theenumi}~}

\section*{Affiliation notes}
\renewcommand\theenumi{\roman{enumi}}
\begin{Authlist}
\item \Adef{0}Deceased
\item \Adef{idp1760560}{Also at: Georgia State University, Atlanta, Georgia, United States}
\item \Adef{idp3121200}{Also at: Also at Department of Applied Physics, Aligarh Muslim University, Aligarh, India}
\item \Adef{idp3827776}{Also at: M.V. Lomonosov Moscow State University, D.V. Skobeltsyn Institute of Nuclear, Physics, Moscow, Russia}
\end{Authlist}

\section*{Collaboration Institutes}
\renewcommand\theenumi{\arabic{enumi}~}
\begin{Authlist}

\item \Idef{org1}A.I. Alikhanyan National Science Laboratory (Yerevan Physics Institute) Foundation, Yerevan, Armenia
\item \Idef{org2}Benem\'{e}rita Universidad Aut\'{o}noma de Puebla, Puebla, Mexico
\item \Idef{org3}Bogolyubov Institute for Theoretical Physics, Kiev, Ukraine
\item \Idef{org4}Bose Institute, Department of Physics and Centre for Astroparticle Physics and Space Science (CAPSS), Kolkata, India
\item \Idef{org5}Budker Institute for Nuclear Physics, Novosibirsk, Russia
\item \Idef{org6}California Polytechnic State University, San Luis Obispo, California, United States
\item \Idef{org7}Central China Normal University, Wuhan, China
\item \Idef{org8}Centre de Calcul de l'IN2P3, Villeurbanne, France
\item \Idef{org9}Centro de Aplicaciones Tecnol\'{o}gicas y Desarrollo Nuclear (CEADEN), Havana, Cuba
\item \Idef{org10}Centro de Investigaciones Energ\'{e}ticas Medioambientales y Tecnol\'{o}gicas (CIEMAT), Madrid, Spain
\item \Idef{org11}Centro de Investigaci\'{o}n y de Estudios Avanzados (CINVESTAV), Mexico City and M\'{e}rida, Mexico
\item \Idef{org12}Centro Fermi - Museo Storico della Fisica e Centro Studi e Ricerche ``Enrico Fermi'', Rome, Italy
\item \Idef{org13}Chicago State University, Chicago, Illinois, USA
\item \Idef{org14}China Institute of Atomic Energy, Beijing, China
\item \Idef{org15}Commissariat \`{a} l'Energie Atomique, IRFU, Saclay, France
\item \Idef{org16}COMSATS Institute of Information Technology (CIIT), Islamabad, Pakistan
\item \Idef{org17}Departamento de F\'{\i}sica de Part\'{\i}culas and IGFAE, Universidad de Santiago de Compostela, Santiago de Compostela, Spain
\item \Idef{org18}Department of Physics and Technology, University of Bergen, Bergen, Norway
\item \Idef{org19}Department of Physics, Aligarh Muslim University, Aligarh, India
\item \Idef{org20}Department of Physics, Ohio State University, Columbus, Ohio, United States
\item \Idef{org21}Department of Physics, Sejong University, Seoul, South Korea
\item \Idef{org22}Department of Physics, University of Oslo, Oslo, Norway
\item \Idef{org23}Dipartimento di Elettrotecnica ed Elettronica del Politecnico, Bari, Italy
\item \Idef{org24}Dipartimento di Fisica dell'Universit\`{a} 'La Sapienza' and Sezione INFN Rome, Italy
\item \Idef{org25}Dipartimento di Fisica dell'Universit\`{a} and Sezione INFN, Cagliari, Italy
\item \Idef{org26}Dipartimento di Fisica dell'Universit\`{a} and Sezione INFN, Trieste, Italy
\item \Idef{org27}Dipartimento di Fisica dell'Universit\`{a} and Sezione INFN, Turin, Italy
\item \Idef{org28}Dipartimento di Fisica e Astronomia dell'Universit\`{a} and Sezione INFN, Bologna, Italy
\item \Idef{org29}Dipartimento di Fisica e Astronomia dell'Universit\`{a} and Sezione INFN, Catania, Italy
\item \Idef{org30}Dipartimento di Fisica e Astronomia dell'Universit\`{a} and Sezione INFN, Padova, Italy
\item \Idef{org31}Dipartimento di Fisica `E.R.~Caianiello' dell'Universit\`{a} and Gruppo Collegato INFN, Salerno, Italy
\item \Idef{org32}Dipartimento di Scienze e Innovazione Tecnologica dell'Universit\`{a} del  Piemonte Orientale and Gruppo Collegato INFN, Alessandria, Italy
\item \Idef{org33}Dipartimento Interateneo di Fisica `M.~Merlin' and Sezione INFN, Bari, Italy
\item \Idef{org34}Division of Experimental High Energy Physics, University of Lund, Lund, Sweden
\item \Idef{org35}Eberhard Karls Universit\"{a}t T\"{u}bingen, T\"{u}bingen, Germany
\item \Idef{org36}European Organization for Nuclear Research (CERN), Geneva, Switzerland
\item \Idef{org37}Excellence Cluster Universe, Technische Universit\"{a}t M\"{u}nchen, Munich, Germany
\item \Idef{org38}Faculty of Engineering, Bergen University College, Bergen, Norway
\item \Idef{org39}Faculty of Mathematics, Physics and Informatics, Comenius University, Bratislava, Slovakia
\item \Idef{org40}Faculty of Nuclear Sciences and Physical Engineering, Czech Technical University in Prague, Prague, Czech Republic
\item \Idef{org41}Faculty of Science, P.J.~\v{S}af\'{a}rik University, Ko\v{s}ice, Slovakia
\item \Idef{org42}Faculty of Technology, Buskerud and Vestfold University College, Vestfold, Norway
\item \Idef{org43}Frankfurt Institute for Advanced Studies, Johann Wolfgang Goethe-Universit\"{a}t Frankfurt, Frankfurt, Germany
\item \Idef{org44}Gangneung-Wonju National University, Gangneung, South Korea
\item \Idef{org45}Gauhati University, Department of Physics, Guwahati, India
\item \Idef{org46}Helsinki Institute of Physics (HIP), Helsinki, Finland
\item \Idef{org47}Hiroshima University, Hiroshima, Japan
\item \Idef{org48}Indian Institute of Technology Bombay (IIT), Mumbai, India
\item \Idef{org49}Indian Institute of Technology Indore, Indore (IITI), India
\item \Idef{org50}Inha University, Incheon, South Korea
\item \Idef{org51}Institut de Physique Nucl\'eaire d'Orsay (IPNO), Universit\'e Paris-Sud, CNRS-IN2P3, Orsay, France
\item \Idef{org52}Institut f\"{u}r Informatik, Johann Wolfgang Goethe-Universit\"{a}t Frankfurt, Frankfurt, Germany
\item \Idef{org53}Institut f\"{u}r Kernphysik, Johann Wolfgang Goethe-Universit\"{a}t Frankfurt, Frankfurt, Germany
\item \Idef{org54}Institut f\"{u}r Kernphysik, Westf\"{a}lische Wilhelms-Universit\"{a}t M\"{u}nster, M\"{u}nster, Germany
\item \Idef{org55}Institut Pluridisciplinaire Hubert Curien (IPHC), Universit\'{e} de Strasbourg, CNRS-IN2P3, Strasbourg, France
\item \Idef{org56}Institute for Nuclear Research, Academy of Sciences, Moscow, Russia
\item \Idef{org57}Institute for Subatomic Physics of Utrecht University, Utrecht, Netherlands
\item \Idef{org58}Institute for Theoretical and Experimental Physics, Moscow, Russia
\item \Idef{org59}Institute of Experimental Physics, Slovak Academy of Sciences, Ko\v{s}ice, Slovakia
\item \Idef{org60}Institute of Physics, Academy of Sciences of the Czech Republic, Prague, Czech Republic
\item \Idef{org61}Institute of Physics, Bhubaneswar, India
\item \Idef{org62}Institute of Space Science (ISS), Bucharest, Romania
\item \Idef{org63}Instituto de Ciencias Nucleares, Universidad Nacional Aut\'{o}noma de M\'{e}xico, Mexico City, Mexico
\item \Idef{org64}Instituto de F\'{\i}sica, Universidad Nacional Aut\'{o}noma de M\'{e}xico, Mexico City, Mexico
\item \Idef{org65}iThemba LABS, National Research Foundation, Somerset West, South Africa
\item \Idef{org66}Joint Institute for Nuclear Research (JINR), Dubna, Russia
\item \Idef{org67}Konkuk University, Seoul, South Korea
\item \Idef{org68}Korea Institute of Science and Technology Information, Daejeon, South Korea
\item \Idef{org69}KTO Karatay University, Konya, Turkey
\item \Idef{org70}Laboratoire de Physique Corpusculaire (LPC), Clermont Universit\'{e}, Universit\'{e} Blaise Pascal, CNRS--IN2P3, Clermont-Ferrand, France
\item \Idef{org71}Laboratoire de Physique Subatomique et de Cosmologie, Universit\'{e} Grenoble-Alpes, CNRS-IN2P3, Grenoble, France
\item \Idef{org72}Laboratori Nazionali di Frascati, INFN, Frascati, Italy
\item \Idef{org73}Laboratori Nazionali di Legnaro, INFN, Legnaro, Italy
\item \Idef{org74}Lawrence Berkeley National Laboratory, Berkeley, California, United States
\item \Idef{org75}Moscow Engineering Physics Institute, Moscow, Russia
\item \Idef{org76}Nagasaki Institute of Applied Science, Nagasaki, Japan
\item \Idef{org77}National Centre for Nuclear Studies, Warsaw, Poland
\item \Idef{org78}National Institute for Physics and Nuclear Engineering, Bucharest, Romania
\item \Idef{org79}National Institute of Science Education and Research, Bhubaneswar, India
\item \Idef{org80}National Research Centre Kurchatov Institute, Moscow, Russia
\item \Idef{org81}Niels Bohr Institute, University of Copenhagen, Copenhagen, Denmark
\item \Idef{org82}Nikhef, Nationaal instituut voor subatomaire fysica, Amsterdam, Netherlands
\item \Idef{org83}Nuclear Physics Group, STFC Daresbury Laboratory, Daresbury, United Kingdom
\item \Idef{org84}Nuclear Physics Institute, Academy of Sciences of the Czech Republic, \v{R}e\v{z} u Prahy, Czech Republic
\item \Idef{org85}Oak Ridge National Laboratory, Oak Ridge, Tennessee, United States
\item \Idef{org86}Petersburg Nuclear Physics Institute, Gatchina, Russia
\item \Idef{org87}Physics Department, Creighton University, Omaha, Nebraska, United States
\item \Idef{org88}Physics Department, Panjab University, Chandigarh, India
\item \Idef{org89}Physics Department, University of Athens, Athens, Greece
\item \Idef{org90}Physics Department, University of Cape Town, Cape Town, South Africa
\item \Idef{org91}Physics Department, University of Jammu, Jammu, India
\item \Idef{org92}Physics Department, University of Rajasthan, Jaipur, India
\item \Idef{org93}Physik Department, Technische Universit\"{a}t M\"{u}nchen, Munich, Germany
\item \Idef{org94}Physikalisches Institut, Ruprecht-Karls-Universit\"{a}t Heidelberg, Heidelberg, Germany
\item \Idef{org95}Purdue University, West Lafayette, Indiana, United States
\item \Idef{org96}Pusan National University, Pusan, South Korea
\item \Idef{org97}Research Division and ExtreMe Matter Institute EMMI, GSI Helmholtzzentrum f\"ur Schwerionenforschung, Darmstadt, Germany
\item \Idef{org98}Rudjer Bo\v{s}kovi\'{c} Institute, Zagreb, Croatia
\item \Idef{org99}Russian Federal Nuclear Center (VNIIEF), Sarov, Russia
\item \Idef{org100}Saha Institute of Nuclear Physics, Kolkata, India
\item \Idef{org101}School of Physics and Astronomy, University of Birmingham, Birmingham, United Kingdom
\item \Idef{org102}Secci\'{o}n F\'{\i}sica, Departamento de Ciencias, Pontificia Universidad Cat\'{o}lica del Per\'{u}, Lima, Peru
\item \Idef{org103}Sezione INFN, Bari, Italy
\item \Idef{org104}Sezione INFN, Bologna, Italy
\item \Idef{org105}Sezione INFN, Cagliari, Italy
\item \Idef{org106}Sezione INFN, Catania, Italy
\item \Idef{org107}Sezione INFN, Padova, Italy
\item \Idef{org108}Sezione INFN, Rome, Italy
\item \Idef{org109}Sezione INFN, Trieste, Italy
\item \Idef{org110}Sezione INFN, Turin, Italy
\item \Idef{org111}SSC IHEP of NRC Kurchatov institute, Protvino, Russia
\item \Idef{org112}Stefan Meyer Institut f\"{u}r Subatomare Physik (SMI), Vienna, Austria
\item \Idef{org113}SUBATECH, Ecole des Mines de Nantes, Universit\'{e} de Nantes, CNRS-IN2P3, Nantes, France
\item \Idef{org114}Suranaree University of Technology, Nakhon Ratchasima, Thailand
\item \Idef{org115}Technical University of Ko\v{s}ice, Ko\v{s}ice, Slovakia
\item \Idef{org116}Technical University of Split FESB, Split, Croatia
\item \Idef{org117}The Henryk Niewodniczanski Institute of Nuclear Physics, Polish Academy of Sciences, Cracow, Poland
\item \Idef{org118}The University of Texas at Austin, Physics Department, Austin, Texas, USA
\item \Idef{org119}Universidad Aut\'{o}noma de Sinaloa, Culiac\'{a}n, Mexico
\item \Idef{org120}Universidade de S\~{a}o Paulo (USP), S\~{a}o Paulo, Brazil
\item \Idef{org121}Universidade Estadual de Campinas (UNICAMP), Campinas, Brazil
\item \Idef{org122}University of Houston, Houston, Texas, United States
\item \Idef{org123}University of Jyv\"{a}skyl\"{a}, Jyv\"{a}skyl\"{a}, Finland
\item \Idef{org124}University of Liverpool, Liverpool, United Kingdom
\item \Idef{org125}University of Tennessee, Knoxville, Tennessee, United States
\item \Idef{org126}University of the Witwatersrand, Johannesburg, South Africa
\item \Idef{org127}University of Tokyo, Tokyo, Japan
\item \Idef{org128}University of Tsukuba, Tsukuba, Japan
\item \Idef{org129}University of Zagreb, Zagreb, Croatia
\item \Idef{org130}Universit\'{e} de Lyon, Universit\'{e} Lyon 1, CNRS/IN2P3, IPN-Lyon, Villeurbanne, France
\item \Idef{org131}V.~Fock Institute for Physics, St. Petersburg State University, St. Petersburg, Russia
\item \Idef{org132}Variable Energy Cyclotron Centre, Kolkata, India
\item \Idef{org133}Warsaw University of Technology, Warsaw, Poland
\item \Idef{org134}Wayne State University, Detroit, Michigan, United States
\item \Idef{org135}Wigner Research Centre for Physics, Hungarian Academy of Sciences, Budapest, Hungary
\item \Idef{org136}Yale University, New Haven, Connecticut, United States
\item \Idef{org137}Yonsei University, Seoul, South Korea
\item \Idef{org138}Zentrum f\"{u}r Technologietransfer und Telekommunikation (ZTT), Fachhochschule Worms, Worms, Germany
\end{Authlist}
\endgroup

\else
\ifbibtex
\bibliographystyle{utphys}
\bibliography{biblio}{}
\else
\input{refpreprint.tex}
\fi
\fi
\else
\iffull
\vspace{0.5cm}
\input{acknowledgements_march2013.tex}
\input{refpaper.tex}
\else
\ifbibtex
\bibliographystyle{utphys}
\bibliography{biblio}{}
\else
\input{refpaper.tex}
\fi
\fi
\fi
\end{document}